\documentclass[aps,prl,reprint,superscriptaddress,nofootinbib]{revtex4-2}
\usepackage{amsmath,amssymb}
\usepackage{graphicx}

\begin{document}
\title{Disformal Closure Selects 3D Scalar–Tensor Gravity without Local Bulk Modes}

\author{M. Shams Nejati}
\email{m\underline{\,\;}shams@iasbs.ac.ir}
\affiliation{School of Physics, Institute for Research in Fundamental
Sciences (IPM),\\ P.O.Box 19395-5531, Tehran, Iran}
\author{M.H.~Vahidinia}
\email{vahidinia@iasbs.ac.ir }
\affiliation{Department of Physics, Institute for Advanced Studies in Basic Sciences (IASBS),
P.O. Box 45137-66731, Zanjan, Iran}

\begin{abstract}
 Three-dimensional Einstein gravity has no local degrees of freedom, yet retains rich global and black-hole physics. Adding a scalar normally spoils this. We show that closure of three-dimensional Horndeski gravity under disformal field redefinitions selects a square-root curvature interaction. A nonlinear analysis finds no local bulk modes on regular branches, while an independent Hamiltonian analysis confirms the result on the timelike-gradient branch. This extends the no-local-bulk-dynamics paradigm beyond pure Einstein gravity and motivates studying its boundary and global sectors.
\end{abstract}

\maketitle

\textit{Introduction.---} Different Lagrangians related by a regular, invertible field redefinition are classically equivalent \cite{Kuntz:2026vhs,Kamefuchi:1961sb}. A family of theories is closed under such transformations if every redefinition maps the action back into the same class. The existence of such closed families is far from obvious and, as we show, can carry nontrivial consequences for the dynamics.

In scalar--tensor gravity, a central class of such transformations is given
by the disformal redefinitions that mix the metric and scalar gradient,
\begin{equation}
  \widetilde g_{\mu\nu}
  =A(\phi,X)g_{\mu\nu}+B(\phi,X)\phi_\mu\phi_\nu,
  \quad
  X=-\frac12\phi_\mu\phi^\mu ,
  \label{eq:disformal-intro}
\end{equation}
where $\phi_\mu=\nabla_\mu\phi$. The closure question is naturally studied within Horndeski theory, the most general scalar--tensor theory with second-order Euler--Lagrange equations, which provides a framework for derivative interactions in modified gravity and cosmology
\cite{Horndeski1974,Deffayet2011,DeffayetSteer2013,Kobayashi2019}. 
In four dimensions, disformal transformations typically map the standard Horndeski theory beyond its original class, thereby generating higher-order derivatives \cite{BettoniLiberati2013,Zumalacarregui:2013pma,Gleyzes:2014qga,Michiwaki:2026xru}. By contrast, the general two-dimensional second-order scalar--tensor family is
closed under generic disformal transformations \cite{ShamsNejatiVahidinia2024}.

This dimensional contrast makes three dimensions the natural setting for classifying which theories are closed under generic disformal transformations.{We therefore focus on three-dimensional Horndeski theory,
whose action is specified by $G_2(\phi,X)$, $G_3(\phi,X)$, and
$G_4(\phi,X)$, see Eq.~\eqref{eq:endmatter-Horndeski-action}.}
\footnote{In three dimensions, the $G_5$ sector can be
absorbed into $G_2,G_3,G_4$ modulo a total divergence;
it does not define an independent interaction. In two dimensions, the $G_5$ sector vanishes identically, while the $G_4$ sector can likewise be absorbed, modulo a total divergence, into the $G_2$ and $G_3$ sectors.\cite{Takahashi:2018yzc}}

Such an exceptional closed sector is particularly significant in three dimensions, where pure Einstein gravity lacks local propagating modes and hence gravitational waves  \cite{Deser:1983tn} yet is an exactly solvable, nontrivial model for studying quantum aspects of gravity \cite{Witten:1988hc}. Adding a generic scalar interaction would ordinarily be expected to introduce a local bulk degree of freedom. Remarkably, the sector selected by disformal closure does not propagate any local bulk modes, extending this feature beyond pure Einstein gravity to a genuine scalar-tensor theory. It therefore provides a new class of models — with a scalar field present at the level of the action — in which the methods and questions developed for three-dimensional Einstein gravity remain fully applicable.

To state this result precisely,  we use ``disformal invariance'' to denote this closure of a family; it need
not be a symmetry of an individual Lagrangian. We use ``topological'' in
the local dynamical sense: no local bulk mode propagates, although global
quantities or boundary degrees of freedom may remain.

We establish this result in three steps. First, disformal closure fixes the Horndeski curvature coupling
$G_4(\phi,X)$ to the unique square-root form $G_4=f(\phi)\sqrt{|X|}$. Second, this coupling admits a decomposition into topological and dynamical pieces. When the weight $f$ is constant, the completed action depends only on the Euler characteristics of the scalar leaves, the two-dimensional surfaces on which $\phi$ is constant; they carry no local propagating mode. When $f$
varies, a nontrivial bulk term appears, but a leaf-shape gauge symmetry
removes the independent deformations of each leaf. Third, a nonlinear
constraint analysis shows that the remaining variables also carry no local
propagating mode and that this conclusion survives the addition of arbitrary
lower-order Horndeski interactions on the regular branches studied below.

\textit{Relation to prior theories.---}
Square-root couplings and nonpropagating scalars occur in cuscuton,
extended cuscuton, and symmetry-based theories
\cite{Afshordi:2006ad,Iyonaga:2018vnu,Tasinato:2020scalarless}.
Chagoya and Tasinato obtain an intrinsic-leaf curvature action from a
singular disformal limit; before adding Einstein gravity, that sector
already has no propagating scalar or tensor modes in unitary gauge
\cite{Chagoya:2018yna}. The disformal stability of DHOST class Ib
\cite{BenAchour:2016cay} and its geometric description with arbitrary
lower-order terms \cite{Langlois:2020xbc} are further direct
structural precedents.

Here we establish closure within three-dimensional Horndeski
under arbitrary regular $A(\phi,X),B(\phi,X)$ and a constructive local
equivalence that class stability alone does not imply. On non-null
patches admitting a regular canonical scalar frame, arbitrary smooth
$G_2,G_3$ with $\Xi\neq0$ can be removed by a regular disformal map to
$S_\phi$. An exact leaf-shape symmetry and the nonlinear constraints
eliminate local bulk modes, with an independent Hamiltonian count on
the regular timelike-gradient, constant-rank branch. The Supplemental
Material gives the correspondence with the earlier geometric action;
boundary, global, and matter-coupling equivalence require separate
analysis.

\textit{Disformal closure and leaf topology.---}
Within Horndeski theory, the relevant curvature interaction is
\begin{equation}
  \begin{split}
    \mathcal L_4={}&G_4(\phi,X)R+G_{4,X}(\phi,X)
    \left[(\Box\phi)^2-\phi_{\mu\nu}\phi^{\mu\nu}\right],
  \end{split}
  \label{eq:quartic-Horndeski-intro}
\end{equation}
where $G_{4,X}\equiv\partial G_4/\partial X$, \(\phi_{\mu\nu}\equiv\nabla_\mu\nabla_\nu\phi\).

We now ask which curvature couplings remain within the Horndeski class after a general regular disformal transformation. Requiring the transformed curvature term and the term proportional to
$G_{4,X}$ to recombine into the Horndeski form in Eq.~\eqref{eq:quartic-Horndeski-intro} gives
\begin{equation}
  \bigl(2XB_{,X}-A_{,X}\bigr)
  \bigl(G_4-2XG_{4,X}\bigr)=0.
  \label{eq:factorized-fourth-family-matching-intro}
\end{equation}
One possibility, $2XB_{,X}-A_{,X}=0$, restricts the allowed
transformations. If closure is required for arbitrary transformations, the other factor must instead vanish,
\begin{equation}
  G_4-2XG_{4,X}=0,
  \label{eq:theory-branch-intro}
\end{equation}
whose real solution on each fixed-sign $X\neq0$ branch is $G_4(\phi,X)=f(\phi)\sqrt{|X|}$. Using the locally invertible scalar redefinition \(\phi=\phi(\Phi)\), chosen such that \(f(\phi(\Phi))|\phi'(\Phi)|=\Phi\), and subsequently relabeling \(\Phi\to\phi\), the selected coupling takes the canonical form
\begin{equation}
  \boxed{G_4(\phi,X)=\phi\sqrt{|X|}.}
  \label{eq:canonical-G4-intro}
\end{equation}

In this canonical frame the fourth-family interaction is
$\mathcal L_4=\phi\mathcal L_{\circ}$, where
\begin{equation}
\mathcal L_{\circ}\equiv
\sqrt{|X|}R + 
\frac{s}{2\sqrt{|X|}}
\left[(\Box\phi)^2-\phi_{\mu\nu}\phi^{\mu\nu}\right],
\quad
s\equiv\operatorname{sgn}X .
\end{equation}
We denote the corresponding scalar-weighted action by
\begin{equation}
S_{\phi}[\phi] = \int_{\mathcal M}d^3x\,\sqrt{|g|}\, \phi\,\mathcal L_{\circ}.
\end{equation}

This scalar redefinition, where regular,also generates correlated lower-family terms $G_3(\phi,X)\Box\phi$ and $G_2(\phi,X)$, whose explicit forms are
given in End Matter.

{
\begin{figure}[t]
\centering
\includegraphics[width=0.7\columnwidth]{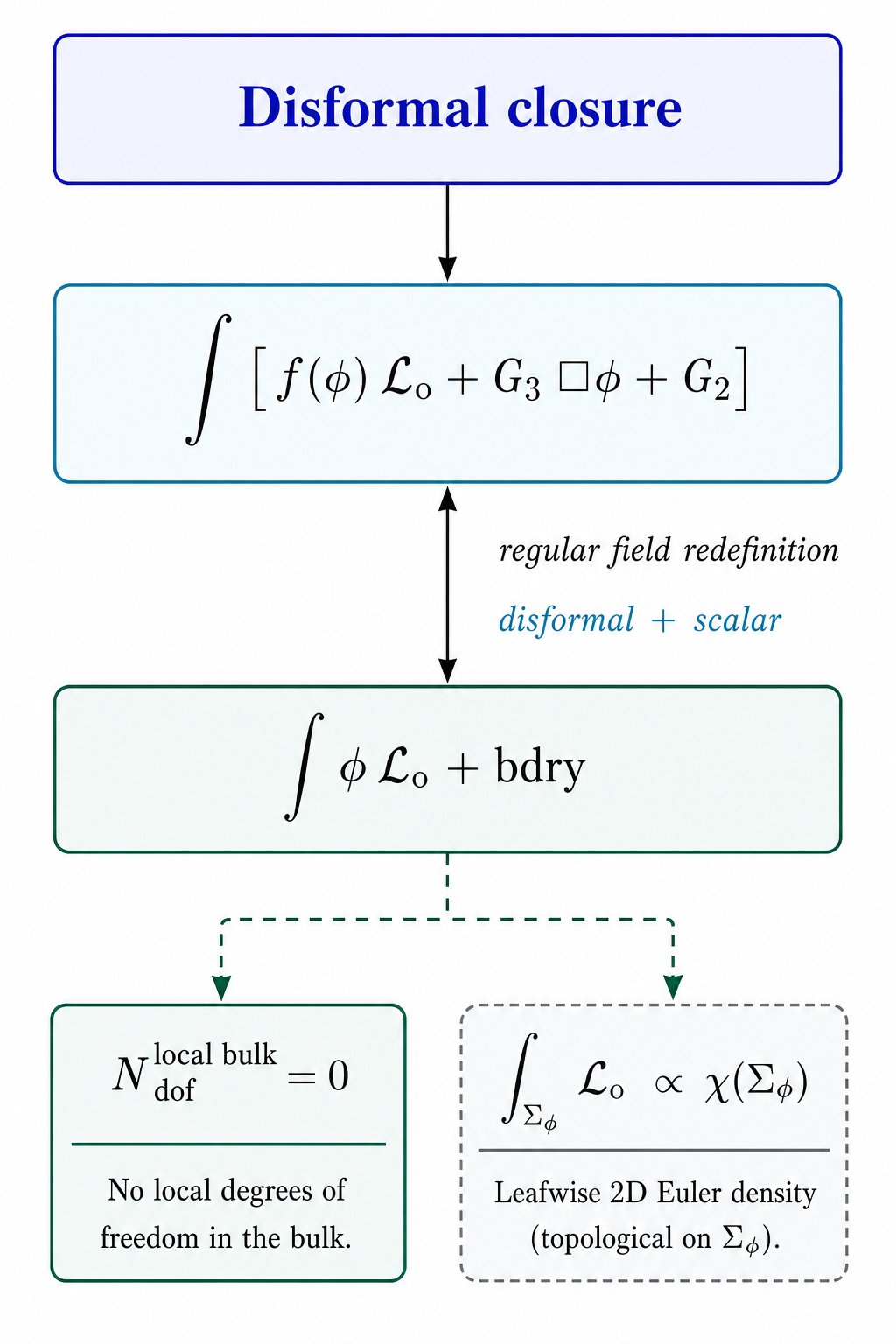}

\caption{
Disformal closure and its relation to the canonical representative.
A regular combination of disformal and scalar-field redefinitions maps the
general theory to the canonical form, while the
\(\mathcal L_{\circ}\) sector has a leafwise topological interpretation.
}
\label{fig:disformal-closure}
\end{figure}
}

The geometric origin of this interaction is particularly simple. On every
smooth fixed-sign patch with $X\neq0$, the level surfaces of $\phi$ foliate
spacetime into two-dimensional leaves
$\Sigma_\phi=\{\phi=\mathrm{const}\}$. The Gauss--Codazzi decomposition gives
the exact identity
\begin{equation}
  \mathcal L_{\circ}
  =
  \sqrt{|X|}\,{\mathcal R}_{\Sigma_{\phi}}+\nabla_\mu J^\mu ,
  \qquad X\neq0,
  \label{eq:leaf-Euler-intro}
\end{equation}
where ${\mathcal R}_{\Sigma_{\phi}}$ is the intrinsic scalar curvature of the leaves and $J^\mu$
is a local current determined by the foliation geometry.

With the Horndeski boundary term \cite{PadillaSivanesan2012} that
completes the leafwise Euler density, the completed unit-weight action
$S_{\circ}^{\rm tot}$ reduces leaf by leaf to the two-dimensional
Gauss--Bonnet invariant,
\begin{equation}
  \boxed{
  S_{\circ}^{\rm tot}
  =
  2\sqrt{2}\,\pi
  \int{\mathrm d}\phi\,\chi(\Sigma_\phi) ,
  }
  \label{eq:Euler-completion-intro}
\end{equation}
depending only on the Euler characteristics of the scalar leaves.
The topological form of $S_{\circ}^{\rm tot}$ also reveals the structural origin of the scalar-weighted theory. Since a constant shift $\phi\to\phi+c$ leaves $X$ and $\mathcal L_{\circ}$ unchanged, the completed action satisfies
\begin{equation}
  S_{\phi}^{\rm tot}[\phi+c] - S_{\phi}^{\rm tot}[\phi] = c\,S_{\circ}^{\rm tot} .
  \label{eq:WZ-shift-law}
\end{equation}
Thus $S_\phi$ realizes the scalar shift in a Wess--Zumino-type manner: it is a scalar-weighted descendant of the topological parent $S_\circ$, rather than an arbitrary scalar--tensor interaction.

\textit{Frozen bulk dynamics.---}
To see how this topological ancestry manifests itself locally, multiply the
leaf-Euler identity, Eq.~\eqref{eq:leaf-Euler-intro}, by $\phi$. Applying
the Leibniz rule to the divergence term gives
\begin{equation}
 \phi\,\mathcal L_{\circ}
 =
 \phi\sqrt{|X|}\,{\mathcal R}_{\Sigma_{\phi}}
 +\nabla_\mu\!\left(\phi J^\mu\right)
 -\phi_\mu J^\mu .
 \label{eq:canonical-transgression}
\end{equation}
The first two terms carry no local bulk dynamics. Because \(\phi\) is constant on each leaf, the integral of the first term over a compact leaf is proportional to \(\phi\,\chi(\Sigma_\phi)\) and contains no derivative transverse to the leaf. The second term is a bulk divergence. Hence the only local, non-topological interaction produced by the varying scalar weight is $-\phi_\mu J^\mu$, which is inherited from the topological parent theory. Its Wess--Zumino-type shift law reflects that origin, and its explicit form shows how the scalar foliation couples neighboring leaves. Indeed, for $X\neq0$, we can define the normal vector $n_\mu=-\phi_\mu/\sqrt{2|X|}$ to write
\begin{equation}
\phi_\mu J^\mu = 2\sqrt2\,|X|K.
 \label{eq:phiJ-K}
\end{equation}
where $K$ denotes the trace of the extrinsic curvature of the leaves and thus measures their normal expansion. The scalar weight therefore couples neighboring leaves through the normal expansion of the foliation. It leads to nontrivial metric equations for $S_{\phi}$, given in Eq.~\eqref{eq:endmatter-EOM-metric} of the End Matter. However, this term is linear in the normal deformation of the leaves and therefore does not generate a nondegenerate kinetic term for the scalar mode. The absence of local propagation is established explicitly from the full nonlinear field equations below.

For the \(S_{\phi}\) theory, the projection of the metric equations along the scalar normal, \(\phi^\mu\mathcal E_{\mu\nu}=0\), gives
\begin{equation}
\nabla_\nu X+\phi_\nu\Box\phi=0.
\end{equation}
Contracting this relation with \(\phi^\nu\) yields
\(\Box\phi=(\phi^\nu\nabla_\nu X)/(2X)\), whereas the trace equation
\(\mathcal E^\mu{}_\mu=0\) gives
\(\Box\phi=-(\phi^\nu\nabla_\nu X)/(2X)\). Therefore, on every regular \(X\neq0\) patch,
\begin{equation}
\nabla_\mu X=0,
\qquad
\Box\phi=0 .
\label{eq:canonical-metric-consequences}
\end{equation}

The scalar Euler equation is not independent: the Noether identity
$2\nabla_\mu\mathcal E^\mu{}_\nu+\mathcal E_\phi\phi_\nu=0$
implies $\mathcal E_\phi=0$ whenever $\mathcal E_{\mu\nu}=0$. At first sight, one may regard $\Box\phi=0$ as a dynamical equation for a scalar mode. However, because it is already a consequence of the metric equations, it is not an independent evolution equation.

This is most clearly seen in scalar-orthogonal coordinates where the metric takes the form
\begin{equation}
 ds^2 =
 -s\,\mathcal N^2 d\phi^2 +h_{ab}dx^a dx^b,\quad
 s=\operatorname{sgn}X ,
 \label{eq:scalar-orthogonal-gauge}
\end{equation}
which can always be chosen locally on a regular non-null patch. Decompose the
leaf metric as
\begin{equation}
 h_{ab} = \rho\,\widehat h_{ab}, \qquad \rho=\sqrt{|h|}, \qquad |\det\widehat h_{ab}|=1 ,
 \label{eq:leaf-area-shape}
\end{equation}
so that $\rho$ is the local leaf volume density and $\widehat h_{ab}$
contains its two local shape components at fixed volume.

In these coordinates, $X=\frac{s}{2\mathcal N^2}$, so the condition $\nabla_\mu X=0$ forces the lapse to be constant, $\mathcal N=\mathcal N_0$, on each connected regular patch. This gives $\Box\phi=-(s/\mathcal N_0^2)\,\partial_\phi\ln\rho$. Requiring $\Box\phi=0$ then yields $\partial_\phi\rho=0$, so the local leaf volume density does not depend on $\phi$: $\rho=\rho(x^a)$.

The remaining dependence of $\rho$ within a leaf is coordinate freedom. The
residual transformations preserving the scalar-orthogonal gauge are
$\phi$-independent leaf diffeomorphisms,
$x^a\rightarrow x'^a(x)$, under which $\rho$ transforms as a
\begin{equation}
\rho'(x')=
\left|\frac{\partial x}{\partial x'}\right|\rho(x)
=\rho_0 ,
\label{eq:rho-local-gauge}
\end{equation}
with $\rho_0$ constant. This is only a local statement: on a closed leaf the
diffeomorphism-invariant volume and $V=\int_{\Sigma_\phi}d^2x\,\rho$
may remain as global data.

Keeping the local leaf volume density fixed, the variation of the leaf metric is
\begin{equation}
\delta h_{ab}
=\delta\rho\,\widehat h_{ab}
+\rho\,\delta\widehat h_{ab}
=\rho\,\delta\widehat h_{ab}
\qquad (\delta\rho=0).
\label{eq:fixed-volume-shape}
\end{equation}
Consequently, fixed-density variations change only the unimodular shape of the leaf metric. The dependence of $K$ on the leaf metric is entirely through $\rho$, not through $\widehat h_{ab}$, since $K=(s/\mathcal N)\partial_\phi\ln\rho$. Moreover, in two dimensions, $\delta(\rho{\mathcal R}_{\Sigma_{\phi}})=\partial_a\Theta^a$. Thus, for compactly supported variations or closed leaves, the local bulk action is invariant under fixed-density deformations of the leaf metric, which are therefore bulk gauge directions.

The metric equations fix $\mathcal N$ and eliminate the normal evolution of $\rho$; residual leaf diffeomorphisms remove its local profile; and fixed-density deformations of $\widehat h_{ab}$ are bulk gauge. Therefore, no local bulk degree of freedom propagates in the theory,
\begin{equation}
\boxed{N_{\rm dof}=0\qquad\text{(local bulk)}}
\end{equation}
This result is fully nonlinear, applies on both fixed-sign branches $X>0$
and $X<0$, and does not rely on perturbation theory or a Hamiltonian degree-of-freedom count. However, we should emphasize that this does not imply that the theory is globally trivial. The global leaf data and possible boundary or edge degrees of freedom must be analyzed separately.

\textit{Local equivalence and the scalarless family.---}
Without loss of generality, we work on a non-null patch admitting the canonical scalar frame
$f(\phi)=\phi$. We construct a local field redefinition relating the
square-root theory with arbitrary smooth $G_2,G_3$ to $S_\phi$
($G_2=G_3=0$). Where this redefinition is invertible, the absence of propagating
modes extends from $S_\phi$ to the corresponding members of the
disformal family.

We seek a regular disformal transformation
\[ \widetilde g_{\mu\nu} = A(\phi,X)g_{\mu\nu} + B(\phi,X)\phi_\mu\phi_\nu \]
for which
\begin{equation}
\sqrt{|\widetilde g|}
\left( \phi\widetilde{\mathcal L}_{\circ} +\widetilde G_3\widetilde\Box\phi +\widetilde G_2
\right) \doteq \sqrt{|g|}\,\phi\mathcal L_{\circ},
\label{eq:canonicalizing-density-identity}
\end{equation}
where \(\doteq\) denotes equality modulo a total divergence.

Writing the transformed kinetic scalar as
$\widetilde X=Y(\phi,X)$, the lower-family matching reduces to
\begin{equation}
\left(
2\sqrt{|Y|}
+Y\widetilde G_{3,Y}
\right)Y_{,\phi}
+
Y\widetilde G_{3,\phi}
+\frac12\widetilde G_2
=0,
\label{eq:canonicalizing-transport}
\end{equation}
where the tilded functions are evaluated at $(\phi,Y)$. Define
\[
\Xi
\equiv
2+s\sqrt{|Y|}\,\widetilde G_{3,Y} .
\]
For $\Xi\neq0$ and a regular local solution with $Y_{,X}\neq0$, the
corresponding disformal functions are
\begin{equation}
\boxed{
A = \frac{2}{\Xi Y_{,X}},
\qquad
B = \frac{1}{\Xi X Y_{,X}} -\frac{1}{2Y}.
}
\label{eq:canonicalizing-disformal-map}
\end{equation}
Thus, $\widetilde G_2$ and $\widetilde G_3$ determine the kinetic
map $Y$, which in turn determines $A,B$. The map is locally invertible
and signature preserving when $\Xi Y_{,X}>0$ and $X/Y>0$.
Suitable initial data ensure these conditions locally near any point
with $Y\neq0$ and $\Xi\neq0$.

The sector $\Xi=0$ must be treated separately. Dropping tildes,
$G_3=-4\sqrt{|X|}+g(\phi)$, and the quartic and cubic terms reduce
to a weighted leaf-Euler interaction with vanishing bulk variation.
With $\widehat G_2=G_2+2Xg_{,\phi}$, the remaining equations are
$\widehat G_2=\widehat G_{2,X}=0$: they select or exclude solution
branches and vanish identically when $\widehat G_2\equiv0$.

Thus, on fixed-sign patches admitting a regular canonical scalar
frame, the regular constant-rank sectors considered here satisfy
\begin{equation*}
\boxed{
G_4=f(\phi)\sqrt{|X|},
\quad
G_2,\;G_3\ \text{arbitrary}
\;\Longrightarrow\;
N_{\rm dof}=0 .
}
\end{equation*}

For example, $G_3=\lambda X$ and $G_2=\eta X^2$ preserve the zero-mode
count for $G_4=\phi\sqrt X$ on $X>0$ with $\lambda>0$. Explicitly, the inverse map to $S_\phi$ has
$A=\Omega^3$ and
$B=\Omega^2(\Omega-1)/(2X)$, where
$\Omega=1-\frac{\sqrt X}{8}[\eta(\phi-\phi_*)+2\lambda\ln(X/X_*)]>0$ for reference constants $\phi_*$ and $X_*>0$.
The Supplemental Material gives the map's derivation,
the reduction and an evolving solution.

The local construction, exceptional-sector reduction, and independent
timelike-branch constraint count are given in the End Matter.

\textit{Candidate global sector.---}
The absence of local bulk propagation does not imply global triviality. To see this, consider the regular branch and assume that $\Sigma_\phi$ has no boundary. We can remove the lapse velocity from the ADM-decomposed Lagrangian, Eq.~\eqref{eq:endmatter-full-ADM}, by adding a boundary term $\Gamma$ defined by $\partial_{\mathcal N}\Gamma(\phi,\mathcal N)\equiv G_3/\mathcal N^{2}$. In this case, the momentum conjugate to the area density $\rho=\sqrt{\det h_{ab}}$ is given by 
$p_\rho=-G_3/\mathcal N-\sqrt2/\mathcal N^2-\Gamma$.
The shift constraint implies that  $p_\rho=p(\phi)$. Leaf diffeomorphisms reduce $\rho$ to the total area $V=\int_{\Sigma_\phi}d^2y\,\rho$, so the canonical pair is described by the symplectic form
\begin{equation}
\Omega_{\rm red}=\delta p\wedge\delta V .
\end{equation}
If the momentum relation is inverted locally as
$\mathcal N=\mathcal N_*(\phi,p)$, define
$\mathcal H=(\Gamma_{,\phi}-\mathcal N G_2)|_{\mathcal N_*}$,
where $\Gamma_{,\phi}$ holds $\mathcal N$ fixed.
Omitting the fixed-topology Euler endpoint term gives the reduced Lagrangian
\begin{equation}
L\doteq 
  p\dot V- H ,
\label{eq:global-reduced-action}
\end{equation}
where $H=V\mathcal H$ is the reduced Hamiltonian.
Here a dot denotes ${\mathrm d}/{\mathrm d}\phi$. Now the dynamical equations are given by Hamilton's equations as
\begin{equation}
\dot V=\partial_{p} H,
\qquad
\dot p=-\partial_{V} H .
\label{eq:global-reduced-evolution}
\end{equation}
Thus, despite the absence of local bulk modes, the candidate global sector can exhibit nontrivial evolution.

\textit{Canonical frame relation.---}
The evolution above has a simple interpretation in the canonical frame.
For $S_\phi$, choosing $G_2=G_3=\Gamma=0$ gives
$H=0$ and $\dot V=\dot p=0$: the candidate global data are constants of motion, although their symplectic pairing need not vanish.

The vanishing of the reduced Hamiltonian is frame-dependent. A regular invertible disformal map relates the generic theory to $S_\phi$ up to a total derivative and preserves the bulk symplectic structure after pullback. Indeed,
\begin{equation}
\mathcal L\to\mathcal L+{\mathrm d}\boldsymbol\mu
\quad\Rightarrow\quad
\boldsymbol\theta\to\boldsymbol\theta+\delta\boldsymbol\mu
\quad\Rightarrow\quad
\widetilde{\boldsymbol\omega}=\boldsymbol\omega,
\end{equation}
where $\boldsymbol\omega=\delta\boldsymbol\theta$, with the equality understood after pullback between the two frames. The regular disformal map therefore induces a canonical transformation on the corresponding scalar-adapted phase space.

With tildes denoting the $S_\phi$ frame, the Hamiltonians are consequently related locally by
\[
H=\widetilde H+\partial_\phi F_2=\partial_\phi F_2,
\]
where $F_2(\widetilde V,p;\phi)$ generates the canonical transformation and $\widetilde H=0$ in the $S_\phi$ frame. Thus the nonvanishing Hamiltonian in a generic frame is generated entirely by the explicit $\phi$ dependence of the canonical transformation.

Consequently, whenever the disformal map is regular across the entire closed leaf, evolving canonical variables in one frame can encode the constant canonical data of $S_\phi$ in the other, yielding a Hamilton--Jacobi-type representation. This pair remains a candidate physical sector pending the final global gauge quotient; spatial boundaries would require a separate completed symplectic analysis.

\textit{BTZ solution.---}
As a nontrivial illustration, consider
\begin{equation}
S = \int d^3x\,\sqrt{|g|}
\left[
\phi\,\mathcal L_{\circ}
+
\frac{2\phi}{\ell^2\sqrt{|X|}}
\left(1+\frac{\phi^2}{\ell^2}\right)
\right].
\label{eq:BTZ-theory}\end{equation}
On the $X<0$ branch, it admits the nonrotating BTZ solution
\begin{equation}
ds^2=-Fdt^2+\frac{dr^2}{F}+r^2d\theta^2,
\quad
\phi=\frac{\ell^{2}\sqrt{F}}{r_+}
\cosh\!\left(\frac{r_+t}{\ell^2}\right),
\label{eq:BTZ-solution}
\end{equation}
where $F(r)=(r^2-r_+^2)/\ell^2$ and $r_+>0$ is the horizon radius.
The scalar extends smoothly across the horizon, and the solution lies in the
$\Xi=2$ sector. Notably, $r_+$ does not enter the action, so the same theory
admits the nonrotating BTZ family with $r_+>0$.

\textit{Dimensional origin.---}
Lovelock boundary terms provide a geometric route to DBI Galileons and
degenerate scalar--tensor interactions
\cite{deRhamTolley2010,ChagoyaTasinato2017}.
Here the completed constant-weight leaf-Euler action arises from a
boundary face of a restricted four-dimensional Gauss--Bonnet construction.
Consider a product geometry with closed, oriented Riemannian leaves
$\Sigma_\phi$ and a Lorentzian two-dimensional factor $P_2$.
A smooth timelike segment $C\subset\partial P_2$ defines the face swept
out by the leaves. Writing $\mathcal R_P$ for the scalar curvature of
$P_2$, the Euler density factorizes as $\mathcal E_4=2{\mathcal R}_{\Sigma_{\phi}}\mathcal R_P$.
Let $k>0$ be the signed extrinsic curvature of $C$, defined using its
outward spacelike normal, and set ${\mathrm d}\phi=4\sqrt2\,k\,{\mathrm d}\tau$,
where $\tau$ is proper time along $C$. The one-form $k\,{\mathrm d}\tau$
measures the normal's turning relative to parallel transport.
The Myers term, the Gauss--Bonnet generalization of the
Gibbons--Hawking--York boundary term, has face density $4k{\mathcal R}_{\Sigma_{\phi}}$;
the boundary-of-face term vanishes in this product geometry
\cite{Myers1987,Cano2018}. The face contribution therefore satisfies
\begin{equation}
\boxed{I_{\rm face}=S_\circ^{\rm tot}.}
\label{eq:GB-parent}
\end{equation}
The four-dimensional bulk, other faces, and genuine joints are
excluded from $I_{\rm face}$. A regular scalar reparametrization
of the seed generates a correlated $G_4+G_3+G_2$ representative
with $\Xi=0$. A separate construction matches the canonical
$\Xi=2$ bulk action by tuning the ambient extrinsic geometry on
a restricted local sector. The Supplemental Material derives both
relations; neither establishes a universal four-dimensional parent.

The leaf geometry also clarifies the dimensional distinction in
disformal closure. A regular disformal transformation conformally
rescales the induced metric on each scalar leaf. In three spacetime
dimensions the leaves are two-dimensional: their curvature integral
changes by a boundary term, canceled by the boundary completion,
leaving the Euler invariant unchanged. In four spacetime dimensions
the leaves are three-dimensional, so this topological cancellation
is unavailable. This does not exclude special four-dimensional
scalar--tensor subclasses.

\textit{Conclusions.---}
Closure under arbitrary regular disformal transformations selects the
square-root Horndeski coupling $G_4=f(\phi)\sqrt{|X|}$.
On regular timelike-gradient patches with $\Xi\neq0$, its leaf-Euler
structure, leaf-shape symmetry, and nonlinear constraints leave no
local bulk modes for arbitrary smooth $G_2,G_3$.
Wherever canonicalization is regular, a constructive scalar--disformal
map relates these theories locally to $S_\phi$; an independent
Hamiltonian count confirms the absence of local modes.
Thus different lower-order interactions share the same local bulk
dynamics on these patches. The separate $\Xi=0$ sector has algebraic
bulk equations on the timelike-gradient branch, which vanish
identically in the weighted-Euler subclass. Boundary completions and
global sectors require separate analysis.

Without local bulk propagation, boundaries and global structure become the natural places to seek physical content. Under relaxed boundary conditions, leaf-shape transformations may carry charges or support edge degrees of freedom; the completed corner symplectic structure is needed to decide which survive as physical data \cite{Donnelly:2016auv,Freidel:2020xyx,Adami:2022ktn}.  On compact closed leaves, the candidate pair $(V,p)$ could supply a finite-dimensional phase space and a possible route to quantization if it survives the full global gauge quotient. Black-hole and cosmological solutions may reveal other global or boundary sectors. The non-null leaf decomposition ceases to apply at $X=0$, where the scalar gradient is null or vanishes. Morse critical points, where the gradient vanishes, offer a setting to investigate whether changes in leaf topology contribute additional structure \cite{Milnor1963}.

A further direction is dimensional. The restricted constructions presented
here connect a four-dimensional Euler invariant to a three-dimensional
foliated theory whose two-dimensional scalar leaves carry their intrinsic
Euler density. This motivates asking whether analogous constructions exist
in odd spacetime dimensions $D=2m+1$, with $2m$-dimensional scalar leaves
carrying the Euler density $\mathcal E_{2m}$. In particular, could such
theories arise as descendants or boundary sectors of the
$(2m+2)$-dimensional Euler invariant $\mathcal E_{2m+2}$, with the
appropriate boundary completion \cite{Myers1987,Cano2018}? Schematically,
\begin{equation}
\mathcal E_{2m+2}
\;\longrightarrow\;
\mathcal L_{\rm fol}^{(2m+1)}
\;\longrightarrow\;
\left.\mathcal E_{2m}\right|_{\Sigma_\phi},
\end{equation}
where $\mathcal E_n$ denotes the $n$-dimensional Euler density and
$\mathcal L_{\rm fol}^{(2m+1)}$ denotes the putative foliated descendant.
For $m=1$, $\mathcal L_{\rm fol}^{(3)}=\mathcal L_\circ$, reproducing the
$4D\!\to3D\!\to2D$ dimensional pattern studied here. For $m>1$, however,
the topological character of the intrinsic leaf functional alone does not
determine the dynamics of the accompanying extrinsic-curvature terms.
Whether such higher-dimensional constructions admit analogous disformal
closure and leaf-shape gauge symmetry, and whether their nonlinear
constraints eliminate all local bulk modes, remains open.

\begin{acknowledgments}
\textit{Acknowledgments.}
We are grateful to M. M. Sheikh-Jabbari and V. Taghiloo for helpful discussions and comments, and to H. Afshar for feedback on the manuscript. The authors acknowledge the use of OpenAI ChatGPT (GPT-5.6 Sol) and, to a limited extent, Anthropic Claude for language editing and clarity, exploratory scientific reasoning, cross-checking derivations, literature analysis, and manuscript preparation. The authors independently verified all calculations, arguments, and references incorporated into the manuscript and take full responsibility for its scientific content.
MSN’s work is based upon research funded by
Iran National Science Foundation (INSF) under project No.4037031.
\end{acknowledgments}

\textit{Data availability.}
No data were created or analyzed in this study.

\bibliography{references}

\appendix

\clearpage
\section*{END MATTER}

The completed three-dimensional Horndeski action is

\begin{align}
S_{\rm H}^{\rm tot}
={}&
\int_{{\mathcal M}}{\mathrm d}^3x\,\sqrt{-g}\,
\Bigl[
G_2+G_3\Box\phi+G_4R
\nonumber\\
&
+G_{4,X}
\left(
(\Box\phi)^2-\phi_{\mu\nu}\phi^{\mu\nu}
\right)
\Bigr]
+S_{\rm H}^{\rm bdy}
+S_{\rm H}^{\rm jnt}.
\label{eq:endmatter-Horndeski-action}
\end{align}
{Here $S_{\rm H}^{\rm bdy}$ is the Horndeski Dirichlet completion
\cite{PadillaSivanesan2012}, given with our boundary conventions in
Eq.~\eqref{eq:endmatter-generalized-GHY} of the Supplemental Material;
$S_{\rm H}^{\rm jnt}$ denotes the joint completion.}

\section{Fourth-family coefficient matching}
With $C\equiv1-2XB/A$ and $A,C>0$, using
$\sqrt{|\widetilde g|}=A^{3/2}\sqrt C\,\sqrt{|g|}$ and
$\widetilde X=X/(AC)$, the Ricci
coefficient is $G_4=\sqrt{AC}\,\widetilde G_4(\phi,\widetilde X)$.
After reducing the third-derivative terms by parts, the transformed fourth
family can be organized as
\begin{equation}
\begin{aligned}
\frac{\sqrt{|\widetilde g|}}{\sqrt{|g|}} \widetilde{\mathcal L}_4
&={}
G_4R +\mathcal F_1\!\left[(\Box\phi)^2-\phi_{\mu\nu}\phi^{\mu\nu}
\right] +\mathcal F_2\mathcal Y\Box\phi
\\
&\hspace{-1cm}+\mathcal F_3(\nabla X)^2
+\mathcal F_4\Box\phi
+\mathcal F_5\mathcal Y^2
+\mathcal F_6\mathcal Y
+\mathcal V
+\nabla_\mu\mathcal W^\mu ,
\end{aligned}
\label{eq:endmatter-G4-reduced}
\end{equation}
where $\mathcal Y\equiv\phi^\mu\nabla_\mu X$. The Horndeski form requires
$\mathcal F_1=G_{4,X}$. With $\Upsilon\equiv G_4-2XG_{4,X}$, the mismatch is
\begin{equation}
\boxed{
\mathcal F_1-G_{4,X}
=
\frac{2XB_{,X}-A_{,X}}
{2\left[AC-X(AC)_{,X}\right]}\,
\Upsilon .
}
\label{eq:endmatter-G4-obstruction}
\end{equation}
The $\mathcal F_i$ are given in the Supplemental Material; in particular,
$\mathcal F_{2,3,5}$ are proportional to $\Upsilon$.

Since $\widetilde X_{,X}=[AC-X(AC)_{,X}]/(AC)^2$,
invertibility requires $AC-X(AC)_{,X}\neq0$.
The mismatch then vanishes either for the restricted transformations
$A_{,X}=2XB_{,X}$ or for $\Upsilon=0$. Closure under arbitrary regular
transformations selects the latter, giving
$G_4=f(\phi)\sqrt{|X|}$ on each fixed-sign $X\neq0$ branch.
The surviving terms reduce, modulo a divergence, to $G_3\Box\phi+G_2$.
Since this lower-family sector is itself disformally closed, the
square-root family is invariant. 

\section{Canonical scalar frame}
A regular scalar redefinition $\phi=\phi(\Phi)$, with
$\mathbf X=-\frac12\Phi_\mu\Phi^\mu$ and $\phi'=d\phi/d\Phi$,
gives $X=\phi'^2\mathbf X$.
The canonical normalization $f(\phi(\Phi))|\phi'|=\Phi$ implies
$\Phi^2(\phi)=c+2\int^\phi f(u)\,du$ for an orientation-preserving map.
The weighted interaction becomes
\begin{equation}
f(\phi)\mathcal L_\circ
=\Phi\mathcal L_\circ+\Delta G_2+\Delta G_3\Box\Phi
+\nabla_\mu V^\mu ,
\end{equation}
where $V^\mu=2\Phi(\phi''/\phi')\sqrt{|\mathbf X|}\,\Phi^\mu$,
$\Delta G_3=-4\Phi(\phi''/\phi')\sqrt{|\mathbf X|}$, and
$\Delta G_2=4\frac{d}{d\Phi}(\Phi\phi''/\phi')\mathbf X\sqrt{|\mathbf X|}$.
Relabeling $\Phi\to\phi$ gives $G_4=\phi\sqrt{|X|}$.

\section{Local construction of the disformal map}
\label{sec:endmatter-local-map}
{In the regular canonical scalar frame, choose initial data
$Y(\phi_0,X)=h(X)$ for Eq.~\eqref{eq:canonicalizing-transport}, with
$X/h>0$ and $\Xi(\phi_0,h(X))h'>0$ near a point where
$h\neq0$ and $\Xi\neq0$. Such data exist locally for either sign of
$\Xi$. Since the coefficient of $Y_{,\phi}$ is
$\sqrt{|Y|}\,\Xi\neq0$, the equation defines a smooth local flow.
Smooth dependence on initial data preserves $Y_{,X}\neq0$, while
continuity preserves $X/Y>0$ and $\Xi Y_{,X}>0$ in a sufficiently
small neighborhood. Substitution into
Eq.~\eqref{eq:canonicalizing-disformal-map} then gives a locally
invertible, signature-preserving map. The Supplemental Material gives
the explicit Jacobian and discusses scalar-frame obstructions.}

\section{Covariant equations for $f(\phi)\mathcal L_\circ$}

For
\(
S_f=\int_{{\mathcal M}}{\mathrm d}^3x\,\sqrt{-g}\,f(\phi)\mathcal L_{\circ}
\),
write the bulk variation as
\(
\delta S_{f}
=\int_{{\mathcal M}}{\mathrm d}^3x\,\sqrt{-g}\,
(\mathcal E_{\mu\nu}\delta g^{\mu\nu}
+\mathcal E_\phi\delta\phi)
+\text{boundary terms}
\).
Direct variation gives
\begin{equation}
\boxed{
\begin{aligned}
\mathcal E_{\mu\nu}=&
\frac{s f_{,\phi}}{\sqrt{|X|}}
\Bigl[
2\phi_{\lambda(\mu}\phi_{\nu)}\phi^\lambda
-g_{\mu\nu}\phi_{\alpha\beta}\phi^\alpha\phi^\beta
-\phi_\mu\phi_\nu\Box\phi
\Bigr]
\\
&-\sqrt{|X|}\,f_{,\phi\phi}
\left(\phi_\mu\phi_\nu+2Xg_{\mu\nu}\right),
\end{aligned}}
\label{eq:endmatter-EOM-metric}
\end{equation}
with symmetrization of weight one half. Constant $f$ gives
$\mathcal E_{\mu\nu}\equiv0$ off shell, while $f=\phi$ gives the
canonical metric equation. Diffeomorphism invariance gives
$\mathcal E_\phi=X^{-1}\phi^\nu\nabla_\mu\mathcal E^\mu{}_\nu$,
so $\mathcal E_{\mu\nu}=0$ also implies $\mathcal E_\phi=0$ for $X\neq0$.

\section{Leafwise boundary completion}

Let \(s=\text{sgn} X\) and \(n_\mu=\sigma\phi_\mu/\sqrt{2|X|}\),
{where $\sigma=\pm1$ selects the normal orientation}; then
\(n_\mu n^\mu=-s\).
Define
\(h_{\mu\nu}=g_{\mu\nu}+s n_\mu n_\nu\),
\(K_{\mu\nu}=h_\mu{}^\rho h_\nu{}^\sigma\nabla_\rho n_\sigma\),
\(K=h^{\mu\nu}K_{\mu\nu}\), and
\(a^\mu=n^\nu\nabla_\nu n^\mu\). On either fixed-sign branch,
the Gauss--Codazzi decomposition gives
\begin{align}
\mathcal L_{\circ}
&=\sqrt{|X|}\,{\mathcal R}_{\Sigma_{\phi}}+\nabla_\mu J^\mu,
\\
J^\mu
&=-2s\sqrt{|X|}\left(a^\mu-Kn^\mu\right).
\label{eq:endmatter-LFE-decomposition}
\end{align}
Multiplying by $\phi$ gives
$\phi\mathcal L_\circ=\phi\sqrt{|X|}{\mathcal R}_{\Sigma_{\phi}}
+\nabla_\mu(\phi J^\mu)-J^\mu\nabla_\mu\phi$.

On the \(X>0\) branch, the Horndeski Dirichlet completion
\cite{PadillaSivanesan2012}, specialized to the constant-weight square-root
theory with roundable joints, cancels the flux of \(J^\mu\) and
supplies the two-dimensional Gibbons--Hawking--York boundary term on each leaf:
\begin{equation}
\begin{aligned}
S_{\circ}^{\rm tot}
=\frac{1}{\sqrt{2}}\int{\mathrm d}\phi\,\biggl[
\int_{\Sigma_\phi}{\mathrm d}^2x\,\sqrt{h}\,{\mathcal R}_{\Sigma_{\phi}}
+2\int_{\partial\Sigma_\phi}{\mathrm d}\ell\,k
\biggr],
\end{aligned}
\label{eq:endmatter-leafwise-completion}
\end{equation}
Here $k$ is the trace of the extrinsic curvature of
$\partial\Sigma_\phi$ within the leaf, using the outward normal.
For compact Riemannian leaves with
smooth boundary, the bracket equals $4\pi\chi(\Sigma_\phi)$, yielding
Eq.~\eqref{eq:Euler-completion-intro}. For roundable corners, Gauss--Bonnet
adds $2\sum_p\vartheta_p$, with signed exterior angles $\vartheta_p$;
Eq.~\eqref{eq:endmatter-leafwise-completion} leaves the corresponding
codimension-two joint term implicit.

\section{Nonlinear local constraint count}

On $X>0$, use $G_4=\phi\sqrt X$ and scalar-adapted ADM coordinates,
\begin{equation}
ds^2=-\mathcal N^2d\phi^2
+h_{ab}(dy^a+\mathcal N^a d\phi)(dy^b+\mathcal N^b d\phi),
\label{eq:endmatter-scalar-ADM}
\end{equation}
with $X=1/(2\mathcal N^2)$.
Write $h_{ab}=\rho\widehat h_{ab}$, $\rho=\sqrt h$,
and $\det\widehat h=1$, leaving two independent shape components.

Modulo a total divergence, the scalar-adapted ADM density of
$G_2+G_3\Box\phi+\phi\mathcal L_\circ$ is

\begin{align}
\mathcal L_{\rm ADM}
&\doteq{}
\frac{\phi}{\sqrt2}\rho\,{\mathcal R}_{\Sigma_{\phi}}+
\rho\frac{G_3}{\mathcal N^2}
\left[
\dot{\mathcal N}
-\mathcal N^a\partial_a\mathcal N
\right] \nonumber\\
&-
\left(
\frac{G_3}{\mathcal N}
+\frac{\sqrt2}{\mathcal N^2}
\right)
\left[
\dot\rho-\partial_a(\rho\mathcal N^a)
\right]+\rho\mathcal N\,G_2.
\label{eq:endmatter-full-ADM}
\end{align}

Only the lapse--area sector has nontrivial first-order kinetic structure;
the traceless shape velocities are absent.
The corresponding momenta obey the primary constraints
\begin{align}
\mathcal C_\rho
&\equiv
p_\rho
+\frac{G_3}{\mathcal N}
+\frac{\sqrt2}{\mathcal N^2}
\approx0,
\label{eq:endmatter-primary-rho}
\\
\mathcal C_{\mathcal N}
&\equiv
p_{\mathcal N}
-\rho\frac{G_3}{\mathcal N^2}
\approx0.
\label{eq:endmatter-primary-lapse}
\end{align}
With $\pi^{ab}$ conjugate to $h_{ab}$, the two independent components of
$P^{ab}\equiv\pi^{ab}-\tfrac12h^{ab}h_{cd}\pi^{cd}\approx0$, together with
the shift momenta $p_a\approx0$, complete the primary set.
The decisive primary bracket is
\begin{align}
\left\{
\mathcal C_{\mathcal N}(x),
\mathcal C_\rho(y)
\right\}
&=
4X^{3/2}\Xi\,
\delta^{(2)}(x-y),
\label{eq:endmatter-primary-bracket}
\end{align}
where, in the canonical scalar frame, $\Xi\equiv2+\sqrt X\,G_{3,X}$.
Thus $(\mathcal C_\rho,\mathcal C_{\mathcal N})$ is a rank-two second-class
pair for $\Xi\neq0$; its rank changes at $\Xi=0$.

The spatial-diffeomorphism constraint obtained from the shift can be written
as
\begin{equation}
\mathcal H_a
=
-2D_bP^b{}_a
-\rho\,\partial_a p_\rho
+p_{\mathcal N}\partial_a\mathcal N
\approx0 .
\label{eq:endmatter-spatial-constraint}
\end{equation}
On the primary surface,
$\mathcal H_a\approx-4\rho X^{3/2}\Xi\,\partial_a\mathcal N$.
Hence $\partial_a\mathcal N\approx0$ for $\Xi\neq0$.
For $G_3=0$, $p_\rho=-\sqrt2/\mathcal N^2$, so
$\partial_a p_\rho\approx0$, as used in the main-text global sector.

The leaf-Euler term has no local shape variation in two dimensions,
so preserving $P^{ab}\approx0$ yields no secondary constraints. The second-class pair
fixes its multipliers, and spatial diffeomorphism invariance closes the
remaining chains.
After eliminating the second-class pair, the shape constraints and the
spatial-diffeomorphism chains are first class with one local reducibility
relation. For an area-preserving vector,
$\partial_a(\rho\xi^a)=0$, the smeared spatial generator satisfies,
{after imposing only the second-class pair and for compactly supported $\xi^a$,}
\begin{equation}
\int_{\Sigma_\phi}{\mathrm d}^2y\,
\xi^a\mathcal H_a
\approx
\int_{\Sigma_\phi}{\mathrm d}^2y\,
P^{ab}\,
2D_{(a}\xi_{b)} .
\end{equation}
The lapse term is $\rho\xi^a\partial_a W(\phi,\mathcal N)$,
where $W_{,\mathcal N}=-4X^{3/2}\Xi$, so its integral vanishes by
$\partial_a(\rho\xi^a)=0$. This establishes reducibility without
imposing $\mathcal H_a=0$.
Since $D_a\xi^a=0$, $2D_{(a}\xi_{b)}$ is traceless:
area-preserving diffeomorphisms are already contained in the shape symmetry.
In two dimensions their single local functional parameter gives one
differential reducibility relation. The six configuration variables
$(h_{ab},\mathcal N,\mathcal N^a)$ give twelve phase-space variables.
The two shape constraints, two shift momenta, and two spatial constraints,
minus one reducibility relation, leave five independent first-class
constraints and the second-class lapse--area pair. Thus
\begin{equation}
\boxed{
N_{\rm dof}
= \frac12 \left[ 12-2(5)-2 \right]=0, \qquad\text{(local bulk)} .
}
\label{eq:endmatter-full-zero-dof}
\end{equation}
This confirms the covariant zero-mode result on the regular
$X>0,\Xi\neq0$ constant-rank branch.
{On an open $\Xi=0$ subclass of either fixed-sign $X\neq0$ branch,}
$G_3=-4\sqrt{|X|}+g(\phi)$ and
$\widehat G_2\equiv G_2+2Xg_{,\phi}$.
{The action takes the form}
\begin{equation}
S_{\Xi=0}\doteq{}\int d^3x\,\sqrt{|g|}\,
\Bigl[\phi\mathcal L_\circ -4\sqrt{|X|}\,\Box\phi+\widehat G_2\Bigr].
\label{eq:Xi-zero-general-action}
\end{equation}
The correlated quartic and cubic terms obey
\begin{equation}
\phi\mathcal L_\circ-4\sqrt{|X|}\,\Box\phi
\doteq\phi\sqrt{|X|}\,\mathcal R_{\Sigma_\phi},
\label{eq:Xi-zero-weighted-Euler}
\end{equation}
{so their local bulk Euler operators vanish identically.
On $X>0$, the lapse--area kinetic term is a divergence and the
scalar-adapted density reduces to}
\begin{equation}
\mathcal L_{\Xi=0}
\doteq
\rho\mathcal N\,\widehat G_2(\phi,X)
+
\frac{\phi}{\sqrt2}\rho\,{\mathcal R}_{\Sigma_{\phi}} ,
\label{eq:endmatter-Xi-zero-reduced}
\end{equation}
with purely algebraic bulk equations
$\widehat G_2=\widehat G_{2,X}=0$.
The generic Dirac count does not extend to $\Xi=0$; isolated or
multiple roots and rank-changing loci require separate constraint analysis.

\clearpage
\onecolumngrid

\setcounter{section}{0}
\setcounter{subsection}{0}
\setcounter{equation}{0}

\setcounter{secnumdepth}{2}

\renewcommand{\thesection}{S\arabic{section}}
\renewcommand{\thesubsection}{\thesection.\arabic{subsection}}
\renewcommand{\theequation}{S\arabic{equation}}

\begin{center}
  \large\bfseries
  Supplemental Material for\\[0.4em]
  ``Disformal Closure Selects 3D Scalar–Tensor Gravity without Local Bulk Modes''
\end{center}

\vspace{1em}

\section{Horndeski action and boundary completion}
\label{sec:supp-boundary-completion}

The three-dimensional Horndeski action, including its boundary and joint
terms, is
\begin{align}
S_{\rm H}^{\rm tot}
={}&\int_{{\mathcal M}}{\mathrm d}^3x\,\sqrt{-g}\,
\Bigl[G_2+G_3\Box\phi+G_4 R \quad
+G_{4,X}\bigl((\Box\phi)^2-\phi_{\mu\nu}\phi^{\mu\nu}\bigr)
\Bigr]
+S_{\rm H}^{\rm bdy}+S_{\rm H}^{\rm jnt}.
\label{eq:endmatter-completed-Horndeski-action}
\end{align}
For each smooth non-null boundary face
$\mathcal B\subset\partial\mathcal M$ (a two-dimensional surface),
let $r_\mu$ be its outward unit normal, $\varepsilon_{\mathcal B}=r_\mu r^\mu=\pm1$,
$\gamma_{ij}$ the induced metric, and $D_i$ its covariant derivative.
The trace of the face's extrinsic curvature is
$\mathcal K=\gamma^{\mu\nu}\nabla_\mu r_\nu$. Split the scalar gradient into
normal and tangential parts by defining
\begin{equation*}
\phi_r\equiv r^\mu\nabla_\mu\phi,\qquad
X_\parallel\equiv-\frac12\gamma^{ij}D_i\phi D_j\phi,\qquad
X=X_\parallel-\frac{\varepsilon_{\mathcal B}}2\phi_r^2.
\end{equation*}
For $n=3,4$, define
\begin{equation}
\mathcal P_n
\equiv
\int_0^{\phi_r}{\mathrm d}\lambda\,
G_n\!\left(
\phi,X_\parallel-\frac{\varepsilon_{\mathcal B}}2\lambda^2
\right).
\label{eq:endmatter-boundary-primitive}
\end{equation}
At fixed $\phi$ and $X_\parallel$, this is an antiderivative with respect
to the normal derivative of the scalar:
$\partial\mathcal P_n/\partial\phi_r=G_n(\phi,X)$.

With the convention $\mathcal L_3=+G_3\Box\phi$, the Dirichlet boundary
term on the smooth faces is
\begin{equation}
\boxed{
S_{\rm H}^{\rm bdy}
=
\sum_{\mathcal B}
\int_{\mathcal B}{\mathrm d}^2x\,\sqrt{|\gamma|}\,
\varepsilon_{\mathcal B}
\left[
-\mathcal P_3
+2G_4\mathcal K
-2\mathcal P_{4,X_\parallel}D^2\phi
\right],
}
\label{eq:endmatter-generalized-GHY}
\end{equation}
where $D^2\phi=\gamma^{ij}D_iD_j\phi$ and
$\mathcal P_{4,X_\parallel}=\partial\mathcal P_4/\partial X_\parallel$.
The factor $\varepsilon_{\mathcal B}$ follows from using the positive
measure $\sqrt{|\gamma|}$ and an outward unit normal. For constant $G_4$
and $G_3=0$, the integrand reduces to the usual
$2\varepsilon_{\mathcal B}G_4\mathcal K$ Gibbons--Hawking--York term
\cite{PadillaSivanesan2012}.

At corners, $S_{\rm H}^{\rm jnt}$ denotes the joint term defined by a
smoothing limit. We consider smooth non-null faces whose pattern of
intersections is preserved under variation and whose corners admit such
a limit. Null faces and joints between faces of different causal types
are outside the present analysis.

\section{Disformal transformation of three-dimensional Horndeski theory}
\label{sec:supp-disformal-3d}

Consider
\begin{equation}
S_{\rm H}^{(3)}=\int {\mathrm d}^3x\,\sqrt{|g|}\left\{G_2+G_3\Box\phi+G_4R+G_{4,X}\left[(\Box\phi)^2-\phi_{\mu\nu}\phi^{\mu\nu}\right]\right\},\quad X=-\frac12\phi_\mu\phi^\mu , \quad  \phi_{\mu\nu}=\nabla_\mu\nabla_\nu\phi,
\label{eq:supp-3d-Horndeski}
\end{equation}
under the scalar--disformal transformation
\begin{equation}
\widetilde g_{\mu\nu}=A(\phi,X)g_{\mu\nu}+B(\phi,X)\phi_\mu\phi_\nu,\qquad C\equiv1-\frac{2XB}{A},
\label{eq:supp-disformal-map}
\end{equation}
on a regular signature-preserving patch, $A>0$, $C>0$. The elementary transformation rules are
\begin{equation}
\widetilde g^{\mu\nu}=\frac1A\left(g^{\mu\nu}-\frac{B}{AC}\phi^\mu\phi^\nu\right),\qquad
\sqrt{|\widetilde g|}=A^{3/2}\sqrt C\,\sqrt{|g|},\qquad
\widetilde X=\frac{X}{AC},
\label{eq:supp-disformal-basic}
\end{equation}
and
\begin{equation}
\widetilde\Box\phi=\frac1{AC}\left[\Box\phi+\frac12\phi^\mu\nabla_\mu\ln\!\left(\frac AC\right)\right].
\label{eq:supp-disformal-box}
\end{equation}

\subsection{Second and third families}
{
The combined second- and third-family sector is closed, modulo a total derivative:
\begin{equation}
\boxed{
\sqrt{|\widetilde g|}\left(\widetilde G_3\widetilde\Box\phi+\widetilde G_2\right)
=
\sqrt{|g|}\left(G_3^{(23)}\Box\phi+G_2^{(23)}\right)
+
\partial_\mu\left(\sqrt{|g|}\,\mathcal H\phi^\mu\right).
}
\label{eq:supp-G23-closure}
\end{equation}
Here
\begin{equation}
G_3^{(23)}
=
\sqrt{\frac AC}\,\widetilde G_3-\mathcal H,
\qquad
G_2^{(23)}
=
A^{3/2}\sqrt C\,\widetilde G_2
-X\sqrt{\frac AC}\,\widetilde G_3\,
\partial_\phi\ln\!\left(\frac AC\right)
+2X\mathcal H_{,\phi},
\label{eq:supp-G23-sector-map}
\end{equation}
with
\begin{equation}
\mathcal H_{,X}
=
\frac12\sqrt{\frac AC}\,\widetilde G_3\,
\partial_X\ln\!\left(\frac AC\right).
\label{eq:supp-H-definition}
\end{equation}
All tilded functions below are understood to be evaluated at \((\phi,\widetilde X)\).
}

\subsection{Fourth family}

We show that the coupling $G_4=f(\phi)\sqrt{|X|}$ retains its quartic
Horndeski form under arbitrary regular scalar--disformal transformations,
up to induced $G_2$ and $G_3$ terms and a total derivative. Using Eq.~\eqref{eq:supp-disformal-basic} and
$\widetilde R = A^{-1}R + \cdots$, the Ricci-scalar coefficient transforms as
\begin{equation}
  G_4(\phi,X) = \sqrt{AC}\,\widetilde G_4(\phi,\widetilde X).
  \label{eq:supp-G4-map}
\end{equation}
We set
\begin{equation}
  \widetilde\Upsilon
  \equiv \widetilde G_4 - 2\widetilde X\,\widetilde G_{4,\widetilde X},
  \qquad
  \mathcal Y \equiv \phi^\mu\nabla_\mu X .
  \label{eq:supp-G4-definitions}
\end{equation}
After reducing third-derivative structures by integration by parts while
retaining the associated current, the transformation reads
\begin{align}
  \frac{\sqrt{|\widetilde g|}}{\sqrt{|g|}}
  \Bigl\{
    \widetilde G_4\widetilde R
    + \widetilde G_{4,\widetilde X}
      \bigl[(\widetilde\Box\phi)^2 - \widetilde\phi_{\mu\nu}\widetilde\phi^{\mu\nu}\bigr]
  \Bigr\}
  ={}&
  G_4 R
  + \mathcal F_1\bigl[(\Box\phi)^2 - \phi_{\mu\nu}\phi^{\mu\nu}\bigr]
  + \mathcal F_2\mathcal Y\Box\phi
  + \mathcal F_3\nabla_\mu X\nabla^\mu X
  \nonumber\\
  &+ \mathcal F_4\Box\phi
  + \mathcal F_5\mathcal Y^2
  + \mathcal F_6\mathcal Y
  + \mathcal V + \nabla_\mu\mathcal W^\mu .
  \label{eq:supp-G4-transform}
\end{align}
The coefficient functions are
\begin{align}
  \mathcal F_1
  &= G_{4,X}
   + \frac{2XB_{,X} - A_{,X}}{2\sqrt{AC}}\,\widetilde\Upsilon ,
  &
  \mathcal F_2
  &= -\frac{B_{,X}}{\sqrt{AC}}\,\widetilde\Upsilon ,
  &
  \mathcal F_3
  &= \frac{A_{,X}^{\,2} - 2(A + XA_{,X})B_{,X}}{2A\sqrt{AC}}\,\widetilde\Upsilon ,
  \nonumber\\
  \mathcal F_4
  &= \frac{2}{\sqrt{AC}}
     \bigl[XB_{,\phi}\widetilde\Upsilon
     + 2BX\,\widetilde G_{4,\phi}
     - A_{,\phi}\widetilde G_4\bigr] ,
  &
  \mathcal F_5
  &= -\frac{A_{,X}B_{,X}}{2A\sqrt{AC}}\,\widetilde\Upsilon ,
  \label{eq:supp-F1-F5}
\end{align}
\begin{equation}
  \mathcal F_6
  = \frac{1}{\sqrt{AC}}
  \left\{
    2(A_{,X} - B)\widetilde G_{4,\phi}
    - \widetilde G_4
      \!\left[
        \frac{A_{,\phi}}{X} + 2A_{,\phi X}
        - 2A_{,\phi}\partial_X\ln(AC)
      \right]
    + \widetilde\Upsilon
      \!\left[
        \frac{A_{,\phi}}{X} - B_{,\phi}
        - A_{,\phi}\partial_X\ln C
      \right]
  \right\} ,
  \label{eq:supp-F6}
\end{equation}
\begin{align}
  \mathcal V
  &= X\sqrt{\frac{A}{C}}
     \left[
       4C\,\widetilde G_4\,
       \partial_\phi\!\left(\tfrac{1}{C}\partial_\phi\ln A\right)
       + \widetilde\Upsilon\,
       \partial_\phi\ln A\,\partial_\phi\ln(AC^2)
     \right] ,
  \label{eq:supp-V}\\
  \mathcal W^\mu
  &= \widetilde G_4\sqrt{AC}
     \left[
       -\partial_X\ln(A^2C)\,\nabla^\mu X
       + \frac{1-C}{XC}\,\phi^\mu\Box\phi
       + \frac{(AB)_{,X}}{A^2C}\,\mathcal Y\phi^\mu
     \right] .
  \label{eq:supp-W}
\end{align}

The four coefficients that obstruct quartic Horndeski closure share the
factor $\widetilde\Upsilon$:
\begin{equation}
  \mathcal F_1 - G_{4,X}
  = \frac{2XB_{,X} - A_{,X}}{2\sqrt{AC}}\,\widetilde\Upsilon ,
  \qquad
  \mathcal F_2,\;\mathcal F_3,\;\mathcal F_5
  \propto \widetilde\Upsilon .
  \label{eq:supp-G4-factorization}
\end{equation}
Setting $\widetilde\Upsilon = 0$ eliminates all four obstructions for any
choice of $A$ and $B$. On a connected fixed-sign branch with
$\widetilde X \neq 0$ this integrates to
\begin{equation}
  \boxed{
    \widetilde G_4(\phi,\widetilde X) = f(\phi)\sqrt{|\widetilde X|}
    \qquad\Longrightarrow\qquad
    G_4(\phi,X) = f(\phi)\sqrt{|X|} .}
  \label{eq:supp-G4-invariant-orbit}
\end{equation}
Introducing
\begin{equation}
  \mathcal L_\circ
  \equiv \sqrt{|X|}\,R
  + \frac{s}{2\sqrt{|X|}}
    \bigl[(\Box\phi)^2 - \phi_{\mu\nu}\phi^{\mu\nu}\bigr] ,
  \qquad s \equiv \text{sgn} X ,
  \label{eq:supp-Lcirc}
\end{equation}
the quartic density is $\mathcal L_4[G_4]=f(\phi)\mathcal L_\circ$.
On the invariant orbit, the coefficient functions satisfy
\begin{equation}
  \mathcal F_1 = G_{4,X} , \qquad
  \mathcal F_2 = \mathcal F_3 = \mathcal F_5 = 0 .
  \label{eq:supp-orbit-coefficients}
\end{equation}
The remaining terms $\mathcal F_4\Box\phi + \mathcal F_6\mathcal Y + \mathcal V$
belong to the lower Horndeski sectors. Introducing $\mathcal I$ locally by
$\mathcal I_{,X} = \mathcal F_6$ and using
\begin{equation}
  \mathcal F_6\mathcal Y
  = -\mathcal I\Box\phi + 2X\mathcal I_{,\phi}
  + \nabla_\mu(\mathcal I\phi^\mu) ,
  \label{eq:supp-F6-reduction}
\end{equation}
the full transformation becomes
\begin{equation}
  \boxed{
    \sqrt{|\widetilde g|}\,f(\phi)\widetilde{\mathcal L}_\circ
    =
    \sqrt{|g|}
    \left[
      f(\phi)\mathcal L_\circ + \Delta G_3\Box\phi + \Delta G_2
    \right]
    + \partial_\mu\!\left[\sqrt{|g|}
      \left(\mathcal W^\mu + \mathcal I\phi^\mu\right)\right] ,}
  \label{eq:supp-G4-orbit-closure}
\end{equation}
with
\begin{equation}
  \Delta G_3 = \mathcal F_4 - \mathcal I ,
  \qquad
  \Delta G_2 = \mathcal V + 2X\mathcal I_{,\phi} .
  \label{eq:supp-G4-descendants}
\end{equation}
The square-root quartic sector is therefore closed under regular
scalar--disformal transformations, with only induced $G_2$ and $G_3$ terms
and a total derivative.

\section{Gauss--Codazzi and Hessian cancellation}

We now show why the quartic term on the square-root branch reduces
to intrinsic curvature on the scalar leaves plus a divergence. On a
connected fixed-sign patch, define
\begin{align}
  s&=\text{sgn} X,
  &n_\mu&=\frac{\sigma\phi_\mu}{\sqrt{2|X|}},
  &h_{\mu\nu}&=g_{\mu\nu}+s n_\mu n_\nu,
  \nonumber\\
  K_{\mu\nu}&=h_\mu{}^\rho h_\nu{}^\sigma\nabla_\rho n_\sigma,
  &K&=h^{\mu\nu}K_{\mu\nu},
  &a^\mu&=n^\nu\nabla_\nu n^\mu .
  \label{eq:supplement-foliation-data}
\end{align}
Here $s$ labels the sign of $X$, $\sigma=\pm1$ fixes the orientation
of the normal, and $\epsilon\equiv n_\mu n^\mu=-s$ is its squared
norm. With $\mathcal N\equiv(2|X|)^{-1/2}$ and
$\dot{\mathcal N}\equiv n^\mu\nabla_\mu\mathcal N$, the scalar
Hessian decomposes as
\begin{equation}
\phi_{\mu\nu}
=\frac{\sigma}{\mathcal N}K_{\mu\nu}
+\frac{\sigma\epsilon}{\mathcal N}
\left(n_\mu a_\nu+n_\nu a_\mu\right)
-\frac{\sigma\epsilon\,\dot{\mathcal N}}{\mathcal N^2}
n_\mu n_\nu .
\label{eq:supplement-Hessian-foliation}
\end{equation}
Its trace and quadratic Horndeski combination are
\begin{align}
\Box\phi
&=\frac{\sigma}{\mathcal N}
\left(K-\frac{\dot{\mathcal N}}{\mathcal N}\right),
\nonumber\\
(\Box\phi)^2-\phi_{\mu\nu}\phi^{\mu\nu}
&=\frac1{\mathcal N^2}
\left(
K^2-K_{\mu\nu}K^{\mu\nu}
-2\epsilon a_\mu a^\mu
-2K\frac{\dot{\mathcal N}}{\mathcal N}
\right).
\label{eq:supplement-Hessian-combination-foliation}
\end{align}
For the square-root coupling $G_4=\sqrt{|X|}$, the quartic
Horndeski density is
\begin{equation}
  \mathcal L_{\circ}
  \equiv\sqrt{|X|}R
  +\frac{s}{2\sqrt{|X|}}
  \left[(\Box\phi)^2-\phi_{\mu\nu}\phi^{\mu\nu}\right].
  \label{eq:supplement-LFE-definition}
\end{equation}
The Gauss--Codazzi identity, with the normal signature made explicit,
reads
\begin{equation}
R={\mathcal R}_{\Sigma_{\phi}}+\epsilon
\left(K^2-K_{\mu\nu}K^{\mu\nu}\right)
+2\epsilon\nabla_\mu\left(a^\mu-Kn^\mu\right).
\label{eq:supplement-Gauss-Codazzi-scalar}
\end{equation}
Substitution into $\mathcal L_\circ$ cancels the terms quadratic in
$K_{\mu\nu}$. The acceleration and $\dot{\mathcal N}$ terms combine
with the Gauss--Codazzi divergence, giving the off-shell identity
\begin{align}
  \mathcal L_{\circ}
  &=\sqrt{|X|}\,{\mathcal R}_{\Sigma_{\phi}}+\nabla_\mu J^\mu,
  \\
  J^\mu
  &=-2s\sqrt{|X|}\left(a^\mu-Kn^\mu\right).
  \label{eq:supplement-LFE-current}
\end{align}
For the canonical scalar weight, the product rule gives
\begin{equation}
  \phi\mathcal L_{\circ}
  =\phi\sqrt{|X|}\,{\mathcal R}_{\Sigma_{\phi}}
  +\nabla_\mu\!\left(\phi J^\mu\right)
  -J^\mu\nabla_\mu\phi.
  \label{eq:supplement-weighted-LFE-decomposition}
\end{equation}
The constant-weight density thus consists of the intrinsic
leaf-curvature term and a divergence. Multiplication by $\phi$
also produces the first-order term $-J^\mu\nabla_\mu\phi$.
These identities hold on both $X>0$ and $X<0$ patches, with the
normal signature encoded by $\epsilon=-s$.

\section{Leaf-shape symmetry and Noether identity}

Up to a divergence, the canonical scalar-weighted bulk action separates two
aspects of the geometry of $\Sigma_\phi$: its intrinsic Euler term and a
first-order descendant proportional to $|X|K$, where $K$ measures the normal
expansion of the leaf area element. This suggests that traceless deformations
of the leaf metric may leave the bulk dynamics unchanged. We now show that
this is an off-shell symmetry.

With the foliation data of Eq.~\eqref{eq:supplement-foliation-data}, consider
the covariant transformation
\begin{equation}
  \delta_\zeta\phi=0,\qquad
  \delta_\zeta g_{\mu\nu}=\zeta_{\mu\nu},\qquad
  n^\mu\zeta_{\mu\nu}=0,\qquad
  h^{\mu\nu}\zeta_{\mu\nu}=0.
  \label{eq:supp-shape-covariant-transformation}
\end{equation}
A two-dimensional leaf metric has one trace component, which controls its
local area element, and two traceless components describing its shape. The
transformation varies only the latter. In scalar-adapted variables,
$\delta_\zeta h_{ab}=\zeta_{ab}$ with $h^{ab}\zeta_{ab}=0$, while the lapse
and shift remain fixed. Consequently,
$\delta_\zeta X=\delta_\zeta\sqrt{|g|}
=\delta_\zeta\sqrt{|h|}=\delta_\zeta n^\mu=0$.

Since $K=|g|^{-1/2}\partial_\mu(\sqrt{|g|}n^\mu)$, these invariances imply
$\delta_\zeta K=0$. The first-order descendant in
Eq.~\eqref{eq:supplement-weighted-LFE-decomposition} is therefore invariant,
consistently with $J^\mu\nabla_\mu\phi=-2\sigma\sqrt2\,|X|K$.

The lower Horndeski sectors are invariant as well. Since
$\delta_\zeta\phi=\delta_\zeta X=0$, any $G_2(\phi,X)$ term is unchanged.
Moreover, both $\sqrt{|g|}$ and $g^{\mu\nu}\phi_\nu$ are unchanged, so the
divergence formula for $\Box\phi$ gives $\delta_\zeta\Box\phi=0$. Terms of
the form $G_3(\phi,X)\Box\phi$ are therefore separately invariant.

It remains to examine the intrinsic Euler term. In scalar-adapted
coordinates $(\phi,y^a)$, its variation is
\begin{align}
  \delta_\zeta S_{\rm intrinsic}
  ={}&\frac{1}{\sqrt2}\int{\mathrm d}\phi\,{\mathrm d}^2y\,\phi
  \Bigl[
  -\sqrt{|h|}\,\mathcal G_{\Sigma_\phi}^{ab}\zeta_{ab}
  +\partial_a\Theta_{\Sigma_\phi}^a[\zeta]
  \Bigr].
  \label{eq:supp-shape-intrinsic-variation}
\end{align}
Because $\Sigma_\phi$ is two-dimensional, its Einstein tensor vanishes
identically, $\mathcal G_{\Sigma_\phi}^{ab}\equiv0$. Hence the intrinsic
Euler term has no bulk variation under $\zeta_{\mu\nu}$; only the displayed
leaf-boundary contribution remains. The canonical bulk action, as well as
its extension by arbitrary $G_2(\phi,X)$ and $G_3(\phi,X)\Box\phi$ terms,
is therefore invariant for compactly supported $\zeta_{\mu\nu}$ or closed
leaves with fixed endpoint data.

For the action under consideration, let $\mathcal E_{\mu\nu}$ denote its
metric Euler derivative. Since $\zeta_{\mu\nu}$ contains two arbitrary
local functions corresponding to traceless leaf-shape deformations, this
off-shell invariance implies the Noether identity
\begin{equation}
  \boxed{
  \left(
  h_\mu{}^\rho h_\nu{}^\sigma
  -\frac12h_{\mu\nu}h^{\rho\sigma}
  \right)
  \mathcal E_{\rho\sigma}\equiv0.}
  \label{eq:supp-shape-Noether-identity}
\end{equation}
Thus the traceless tangential projection of the metric field equation
vanishes identically and does not constitute an independent equation of
motion. For $S_\phi$, the remaining normal and trace projections yield
Eq.~\eqref{eq:canonical-metric-consequences}; diffeomorphism invariance
then makes the scalar equation dependent on the metric equations on a
regular $X\neq0$ patch.

This establishes a local bulk gauge symmetry for compactly supported
transformations. On closed leaves, the displayed leaf-boundary term also
vanishes. Whether transformations reaching a boundary carry charges
depends on the boundary conditions and the completed boundary and corner
phase space.

\section{Disformal removal of the generic \(G_2\) and \(G_3\) terms}
On the square-root branch with \(f(\phi)=\phi\), we seek a regular
scalar--disformal map that removes arbitrary smooth
\(\widetilde G_2\) and \(\widetilde G_3\). In this section, tildes label
the interacting theory, while untilded variables label the canonical
theory \(S_\phi\). We will show that 
\begin{equation}
\boxed{
\sqrt{|\widetilde g|}
\left(\phi\widetilde{\mathcal L}_{\circ}
+\widetilde G_3\widetilde\Box\phi+\widetilde G_2\right)
=
\sqrt{|g|}\,\phi\mathcal L_{\circ}
+\partial_\mu\left(\sqrt{|g|}\,\mathcal J^\mu\right).
}
\label{eq:supp-disformal-removal-goal}
\end{equation}

Combining Eqs.~\eqref{eq:supp-G23-closure},
\eqref{eq:supp-G4-orbit-closure}, and
\eqref{eq:supp-G4-descendants} gives
\begin{equation}
\frac{\sqrt{|\widetilde g|}}{\sqrt{|g|}}
\left(\phi\widetilde{\mathcal L}_{\circ}
+\widetilde G_3\widetilde\Box\phi+\widetilde G_2\right)
=
\phi\mathcal L_{\circ}
+G_3\Box\phi+G_2
+\nabla_\mu\left[
\mathcal W^\mu+(\mathcal I+\mathcal H)\phi^\mu
\right],
\label{eq:supp-full-orbit-reduced}
\end{equation}
where
\begin{equation}
G_3=\Delta G_3+G_3^{(23)},\qquad
G_2=\Delta G_2+G_2^{(23)}.
\label{eq:supp-G23-total}
\end{equation}
We require both untilded coefficients to vanish. From \(G_3=0\),
\begin{equation}
\mathcal I+\mathcal H
=\mathcal F_4+\sqrt{\frac AC}\,\widetilde G_3,
\qquad
\mathcal F_6
=\mathcal F_{4,X}
+\sqrt{\frac AC}\,
\widetilde G_{3,\widetilde X}\,\widetilde X_{,X}.
\label{eq:supp-G3-chain}
\end{equation}
On \(\widetilde G_4=\phi\sqrt{|\widetilde X|}\), the coefficients also give
\begin{equation}
\mathcal F_{4,X}-\mathcal F_6
=
\frac{2\sqrt{|X|}}{C}
\left[\frac{1-C}{X}-\partial_X\ln(AC)\right].
\end{equation}
Using \(\widetilde X=X/(AC)\), these relations reduce to
\begin{equation}
\widetilde X_{,X}
=\frac{2C\,\widetilde X}{\Xi X},
\qquad
\Xi\equiv
2+s\sqrt{|\widetilde X|}\,
\widetilde G_{3,\widetilde X}.
\label{eq:supp-Xtilde-X}
\end{equation}

The condition \(G_2=0\) fixes the remaining dependence of
\(\widetilde X\). Taking the \(\phi\)-derivative of the first relation
in Eq.~\eqref{eq:supp-G3-chain} and using the coefficient formulas yields
\begin{equation}
\mathcal V+2X\mathcal F_{4,\phi}
=-\frac{4X\sqrt{|X|}}{C}\,\partial_\phi\ln(AC),
\end{equation}
and hence
\begin{equation}
\boxed{
\widetilde X_{,\phi}
=
-\frac{\widetilde G_2
+2\widetilde X\,\widetilde G_{3,\phi}}
{2\sqrt{|\widetilde X|}\,\Xi}.
}
\label{eq:supp-Xtilde-flow}
\end{equation}
The tilded functions in this equation are evaluated at
\((\phi,\widetilde X)\).

At fixed $X$, Eq.~\eqref{eq:supp-Xtilde-flow} is a first-order
ordinary differential equation in $\phi$ for $\widetilde X$.
An initial profile $\widetilde X(\phi_0,X)=h(X)$ selects a local
solution. Equation~\eqref{eq:supp-Xtilde-X} then fixes $C$, and
$AC=X/\widetilde X$ gives
\begin{equation}
\boxed{
A=\frac{2}{\Xi\,\widetilde X_{,X}},
\qquad
B=\frac{1}{\Xi X\,\widetilde X_{,X}}
-\frac{1}{2\widetilde X}.
}
\label{eq:supp-AB-reconstruction}
\end{equation}
By construction, $G_2=G_3=0$, so
Eq.~\eqref{eq:supp-full-orbit-reduced} reduces to
Eq.~\eqref{eq:supp-disformal-removal-goal}, with
\begin{equation}
\mathcal J^\mu
=
\mathcal W^\mu
+\left(\mathcal F_4+\sqrt{\frac AC}\,\widetilde G_3\right)\phi^\mu.
\label{eq:supp-disformal-current}
\end{equation}

The construction may therefore be summarized schematically as
\begin{equation}
(\widetilde G_2,\widetilde G_3)
\;\xrightarrow{\;\text{solve for }\widetilde X\;}\;
\widetilde X(\phi,X)
\;\xrightarrow{\;\widetilde X_{,X}\;}\;
C(\phi,X)
\;\xrightarrow{\;AC=X/\widetilde X\;}\;
A(\phi,X),\,B(\phi,X).
\label{eq:supp-disformal-removal-summary}
\end{equation}
Different admissible profiles $h(X)$ give different maps removing
the same $\widetilde G_2$ and $\widetilde G_3$.

On a fixed-sign branch, signature preservation and local invertibility
require
\begin{equation}
\frac{X}{\widetilde X}>0,\qquad
\Xi\,\widetilde X_{,X}>0.
\label{eq:supp-disformal-regularity}
\end{equation}
These inequalities give $A>0$, $C>0$, and
$\widetilde X_{,X}\neq0$.

\subsection{Local existence and invertibility}
\label{sec:supp-normal-form-existence}

The matching equations determine a candidate disformal map. To relate
the two theories, we must also show that the map preserves the metric
signature and has a local inverse. The first-order equation for
$\widetilde X$ leaves its initial profile $h(X)$ free; we show that this
freedom can meet both requirements for arbitrary smooth
$\widetilde G_2$ and $\widetilde G_3$ near any point where
$\widetilde X\neq0$ and $\Xi\neq0$.

Let $(\phi_0,Y_0)$ be such a point. Choose $X_0$ with the same sign
as $Y_0$, and a smooth profile $h(X)$ such that
$h(X_0)=Y_0$ and $h'(X_0)\Xi(\phi_0,Y_0)>0$.
A linear profile with the required slope provides such initial data.
The right-hand side of Eq.~\eqref{eq:supp-Xtilde-flow} is smooth near
this point, so the equation has a local solution
$\widetilde X(\phi,X)$ that depends smoothly on the initial profile.
Differentiating that equation with respect to $X$ gives a homogeneous
linear equation for $\widetilde X_{,X}$. Consequently,
$\widetilde X_{,X}$ retains the sign of $h'$. On a sufficiently small
neighborhood, $X$ and $\widetilde X$ retain the same sign, while
$\Xi$ retains its initial sign. The reconstructed map therefore
satisfies
\begin{equation}
A=\frac{2}{\Xi\,\widetilde X_{,X}}>0,\qquad
A-2XB=\frac{X}{\widetilde X}>0,\qquad
C=\frac{\Xi X\widetilde X_{,X}}{2\widetilde X}>0.
\label{eq:supp-normal-form-signs}
\end{equation}
Thus a regular, locally invertible map exists near any such point
for smooth $\widetilde G_2$ and $\widetilde G_3$, on either
fixed-sign branch. If $\Xi>0$ initially, one may choose
$X_0=Y_0$ and $h(X)=X$. For $\Xi<0$, one chooses a decreasing
initial profile.

This result assumes that the scalar weight can first be put in
canonical form through a smooth, invertible redefinition satisfying
$f(\phi(\Phi))|{\mathrm d}\phi/{\mathrm d}\Phi|=\Phi$. Such a redefinition is
available locally where $f\neq0$, but is obstructed at
$f=f_{,\phi}=0$. The map is local and does not extend by this
argument across $X=0$, $\Xi=0$, or a loss of smoothness.
Boundary comparisons must include the divergence in
Eq.~\eqref{eq:supp-disformal-removal-goal}; any matter action must
be transformed with the scalar--tensor action.

\section{An interacting example on the regular branch}
\label{sec:supp-interacting-example}

Consider the theory
\begin{equation}
S[g,\phi]=\int {\mathrm d}^3x\,\sqrt{-g}\,
\left[\phi\mathcal L_\circ
+\lambda X\Box\phi+\eta X^2\right],
\label{eq:supp-example-explicit-action}
\end{equation}
where $\mathcal L_\circ$ is the square-root quartic density in
Eq.~\eqref{eq:supp-Lcirc}, $\lambda>0$, and $\eta\neq0$.
Thus $G_4=\phi\sqrt X$, $G_3=\lambda X$, and $G_2=\eta X^2$.

On $X>0$, use the scalar-adapted coordinates and ADM variables of
Eq.~\eqref{eq:endmatter-scalar-ADM}. In particular,
$X=1/(2\mathcal N^2)$ and
$\rho=\sqrt{\det h_{ab}}$ is the area density of a scalar leaf.
Here a dot denotes $\partial_\phi$ at fixed $y^a$. Integrating the
lapse velocity in Eq.~\eqref{eq:endmatter-full-ADM} by parts gives
\begin{equation}
\mathcal L_{\rm ADM}\doteq
p_\rho(\mathcal N)
\left[\dot\rho-\partial_a(\rho\mathcal N^a)\right]
+\frac{\eta\rho}{4\mathcal N^3}
+\frac{\phi}{\sqrt2}\rho{\mathcal R}_{\Sigma_{\phi}} ,
\end{equation}
with area momentum
\begin{equation}
p_\rho(\mathcal N)
=-\frac{\sqrt2}{\mathcal N^2}
-\frac{\lambda}{3\mathcal N^3}.
\end{equation}
Its derivative is
\begin{equation}
\frac{\partial p_\rho}{\partial\mathcal N}
=\frac{2\sqrt2}{\mathcal N^3}
+\frac{\lambda}{\mathcal N^4}
=4X^{3/2}\Xi>0,
\qquad
\Xi=2+\lambda\sqrt X.
\end{equation}
Hence the momentum relation determines the lapse locally as
$\mathcal N=\mathcal N(p_\rho)$ throughout this branch.

For closed leaves of fixed topology, the intrinsic-curvature
integral contributes only a fixed endpoint term. The Hamiltonian
per unit area defined in the Main Text becomes
$\mathcal H(p_\rho)
=-\eta/[4\mathcal N(p_\rho)^3]$.
Integrating the shift term by parts, the remaining bulk density is
\begin{equation}
\mathcal L_{\rm bulk}\doteq
p_\rho\dot\rho-\rho\mathcal H(p_\rho)
+\rho\mathcal N^a\partial_a p_\rho .
\end{equation}
Variation of the shift gives $\partial_a p_\rho=0$, without
assuming spatial homogeneity. The other equations then read
\begin{equation}
\dot p_\rho=-\mathcal H(p_\rho),\qquad
\dot\rho-\partial_a(\rho\mathcal N^a)
=\rho\,\mathcal H_{,p_\rho}.
\end{equation}
Thus $p_\rho=p(\phi)$ and $\mathcal N=\mathcal N(\phi)$ on each
connected leaf. Spatial diffeomorphisms remove local variations
of the area density, while the leaf-shape symmetry removes the
traceless metric variables. Both interactions therefore leave
the local bulk degree-of-freedom count at zero.

The remaining variables can nevertheless evolve. For a flat
torus with vanishing shift and uniform $\rho$, the equations give
\begin{equation}
\dot{\mathcal N}
=\frac{\eta\mathcal N}
 {4(2\sqrt2\mathcal N+\lambda)},
\qquad
\frac{\dot\rho}{\rho}
=\frac{3\eta}
 {4(2\sqrt2\mathcal N+\lambda)}.
\end{equation}
With $\mathcal N(\phi_0)=\mathcal N_0>0$ and
$\rho(\phi_0)=\rho_0>0$, the solution is
\begin{equation}
2\sqrt2(\mathcal N-\mathcal N_0)
+\lambda\ln\frac{\mathcal N}{\mathcal N_0}
=\frac{\eta}{4}(\phi-\phi_0),
\qquad
\rho=\rho_0
\left(\frac{\mathcal N}{\mathcal N_0}\right)^3.
\end{equation}
It applies wherever $\mathcal N>0$. This evolution does not
establish whether its global integration constants survive the
full gauge quotient.

\subsection{Explicit map to the canonical representative}
\label{sec:supp-example-canonical-map}

We now map the interacting metric $g_{\mu\nu}$ to the metric
$\widetilde g_{\mu\nu}$ of $S_\phi$, keeping the scalar fixed:
\begin{equation}
\widetilde g_{\mu\nu}
=A(\phi,X)g_{\mu\nu}+B(\phi,X)\phi_\mu\phi_\nu,
\qquad
S_\phi[\widetilde g,\phi]
=\int{\mathrm d}^3x\,\sqrt{-\widetilde g}\,
\phi\widetilde{\mathcal L}_\circ .
\label{eq:supp-example-target-action}
\end{equation}
Thus untilded quantities describe the interacting theory and
tilded quantities describe the canonical representative. With
this frame assignment, the inverse kinetic relation
$\widetilde X(\phi,X)$ obtained from
Eq.~\eqref{eq:canonicalizing-transport} obeys
\begin{equation}
\widetilde X_{,\phi}
=\frac{\eta X^{3/2}}{2\Xi}\,
\widetilde X_{,X}.
\label{eq:supp-example-inverse-transport}
\end{equation}
Along a curve of constant $\widetilde X$, this equation fixes
how $X$ changes with $\phi$. One convenient local solution is
\begin{equation}
\widetilde X=\frac{X}{\Omega^2},
\qquad
\Omega=1-\frac{\sqrt X}{8}
\left[
\eta(\phi-\phi_*)
+2\lambda\ln\!\left(\frac{X}{X_*}\right)
\right],
\label{eq:supp-example-kinetic-map}
\end{equation}
where $X_*>0$ and $\phi_*$ select the map; they are not new
couplings. Direct differentiation gives
$\widetilde X_{,X}=\Xi/(2\Omega^3)$ and
$\widetilde X_{,\phi}=\eta X^{3/2}/(4\Omega^3)$,
verifying Eq.~\eqref{eq:supp-example-inverse-transport}.
The disformal coefficients are therefore
\begin{equation}
A=\frac{\Xi}{2\widetilde X_{,X}}=\Omega^3,
\qquad
B=\frac{A}{2X}-\frac{1}{2\widetilde X}
=\frac{\Omega^2(\Omega-1)}{2X}.
\label{eq:supp-example-explicit-map}
\end{equation}
They give the action-density relation, modulo a total divergence,
\begin{equation}
\sqrt{-g}\left[
\phi\mathcal L_\circ+\lambda X\Box\phi+\eta X^2
\right]
\doteq
\sqrt{-\widetilde g}\,
\phi\widetilde{\mathcal L}_\circ .
\label{eq:supp-example-density-map}
\end{equation}
On the patch $\Omega>0$,
\begin{equation}
A=\Omega^3>0,\qquad
A-2XB=\Omega^2>0,\qquad
\widetilde X_{,X}=\frac{\Xi}{2\Omega^3}>0.
\label{eq:supp-example-map-regularity}
\end{equation}
Because $\Omega(\phi_*,X_*)=1$, this regular patch exists
around each chosen reference point. The map is local and becomes
singular at $\Omega=0$. The bulk density relation alone does not
identify boundary completions or global sectors.

\section{Comparison with related degenerate theories}
\label{sec:supp-related-degenerate-theories}

The square-root quartic coupling satisfies the class-Ib conditions of
Ref.~\cite{BenAchour:2016cay}, while
Ref.~\cite{Langlois:2020xbc} provides a geometric action for that
class. This geometry is especially useful
in three dimensions: the constant-$\phi$ surfaces are two-dimensional,
and their completed intrinsic-curvature action is the Euler invariant.
We first make the geometric correspondence explicit, then distinguish
it from the disformal, symmetry, and constraint results established
here.

\subsection{Correspondence with the geometric class-I action}

Langlois, Noui, and Roussille express class-I theories through the
intrinsic curvature and extrinsic curvature of constant-$\phi$
surfaces; their class-Ib representative has no Einstein--Hilbert term
\cite{Langlois:2020xbc}. Specializing their geometric form and
lower-order extension to three spacetime dimensions gives
\begin{equation}
S_{\rm geom}^{(3)}
=\int{\mathrm d}^3x\,\sqrt{|g|}
\left[\lambda_1(\phi,Y){\mathcal R}_{\Sigma_{\phi}}
+\nu_1(\phi,Y)K+\nu_0(\phi,Y)\right],
\label{eq:supp-LNR-geometric-action}
\end{equation}
where $K$ measures the normal expansion of the scalar leaves. In this
comparison, $Y=g^{\mu\nu}\phi_\mu\phi_\nu=-2X$,
$q=\sqrt{|Y|}$, and the oriented normal is $n_\mu=\phi_\mu/q$.

For the square-root coupling, choose
\begin{equation}
\lambda_1=\frac{f(\phi)q}{\sqrt2}
=f(\phi)\sqrt{|X|}.
\label{eq:supp-LNR-quartic-map}
\end{equation}
In scalar-adapted coordinates,
$\sqrt{|g|}=\sqrt{|h|}/q$, so
$\sqrt{|g|}\lambda_1{\mathcal R}_{\Sigma_{\phi}}
=f(\phi)\sqrt{|h|}{\mathcal R}_{\Sigma_{\phi}}/\sqrt2$.
The coefficient is constant over each leaf; on closed leaves, its
intrinsic-curvature integral depends only on the leaf topology.
A varying $f(\phi)$ nevertheless leaves a term proportional to the
leaf expansion:
\begin{equation}
f\mathcal L_\circ
\doteq \lambda_1{\mathcal R}_{\Sigma_{\phi}}+\sqrt2\,|Y|f_{,\phi}K .
\label{eq:supp-LNR-weighted-map}
\end{equation}
Here $\doteq$ denotes equality up to a divergence. Equation
\eqref{eq:supp-LNR-weighted-map} follows from the foliation current
in Eq.~\eqref{eq:supplement-LFE-current} with the normal orientation
specified above.

The same geometric action accommodates arbitrary lower-order terms.
To express $G_3\Box\phi$ through $K$, introduce a function $H(\phi,Y)$
that solves, locally on $Y\neq0$,
\begin{equation}
H+2YH_{,Y}=-2G_3(\phi,-Y/2).
\label{eq:supp-LNR-primitive}
\end{equation}
The coefficients that reproduce
$f\mathcal L_\circ+G_3\Box\phi+G_2$ up to a divergence are
\begin{equation}
\nu_1=\sqrt2\,|Y|f_{,\phi}-YqH_{,Y},
\qquad
\nu_0=G_2(\phi,-Y/2)+\frac{Y}{2}H_{,\phi}.
\label{eq:supp-LNR-lower-map}
\end{equation}
The key step in checking this relation is to integrate by parts the
$\phi^\mu\phi^\nu\phi_{\mu\nu}$ term:
\begin{equation}
H_{,Y}\phi^\mu\phi^\nu\phi_{\mu\nu}
\doteq-\frac12H\Box\phi-\frac{Y}{2}H_{,\phi}.
\label{eq:supp-LNR-IBP}
\end{equation}
Using
$K=(\Box\phi-\phi^\mu\phi^\nu\phi_{\mu\nu}/Y)/q$
then gives
\begin{equation}
S_{\rm geom}^{(3)}
\doteq\int{\mathrm d}^3x\,\sqrt{|g|}
\left[f\mathcal L_\circ+G_3\Box\phi+G_2\right].
\label{eq:supp-LNR-full-correspondence}
\end{equation}
Thus the inclusion of arbitrary $G_2$ and $G_3$ in this geometric
structure is an earlier result. The equations above give its explicit
three-dimensional correspondence, including the descendant of the
weighted divergence.

The coefficient classification provides a related, but narrower,
comparison. In the conventions of Ref.~\cite{BenAchour:2016cay},
let $F(\phi,Y)=G_4(\phi,-Y/2)$. The quartic Horndeski coefficients obey
\begin{equation}
\alpha_1=-\alpha_2=2F_{,Y}=-G_{4,X},
\qquad
\alpha_3=\alpha_4=\alpha_5=0 .
\end{equation}
For $G_4=f(\phi)\sqrt{|X|}$,
\begin{equation}
F-Y\alpha_1=G_4-2XG_{4,X}=0,
\end{equation}
which is the class-Ib coefficient condition. That class is stable
under invertible disformal transformations
\cite{BenAchour:2016cay}. This four-dimensional coefficient
classification does not determine the local mode count in three
dimensions or establish closure within three-dimensional Horndeski
theory. Relatedly, Ref.~\cite{Katayama:2025hnd} considers closure under
$A(\phi),B(\phi)$ transformations, whereas our closure test allows
arbitrary regular $X$ dependence.

Extended cuscuton offers another comparison on the $X>0$ branch.
In the ADM convention where $A_4$ multiplies
$K^2-K_{ab}K^{ab}$, quartic Horndeski gives
$A_4=-G_4+2XG_{4,X}=0$ for our family. For the extended-cuscuton
Horndeski form
$G_4=b_0(\phi)+b_1(\phi)\sqrt{2X}$, one instead has
$A_4=-b_0$. The usual Hamiltonian analysis of its $A_5=0$ branch
assumes $A_4\neq0$ \cite{Iyonaga:2018vnu}. Setting $b_0=0$ removes
the quadratic metric-velocity block used in that analysis, so its
mode count cannot simply be carried over to this different-rank
sector. This observation does not rule out singular limits or other
extensions.

A fourth comparison is with symmetry-based theories having no
propagating scalar mode
\cite{Chagoya:2018yna,Tasinato:2020scalarless}. Their metric
transformation acts as
\begin{equation}
\delta_\epsilon g_{\mu\nu}
=2\phi_{(\mu}\nabla_{\nu)}\epsilon,
\qquad
\delta_\epsilon\phi=0.
\end{equation}
Writing $u=\phi^\mu\nabla_\mu\epsilon$, its action on a lower-order
potential is
\begin{equation}
\delta_\epsilon\!\left(\sqrt{|g|}\,G_2\right)
=\sqrt{|g|}\left(G_2-2XG_{2,X}\right)u.
\end{equation}
For example, $G_2=\eta X^2$ gives
$-3\eta\sqrt{|g|}\,X^2u$; the symmetry therefore does not survive
a generic $G_2$ term. By contrast, the traceless tangential
transformation of Eq.~\eqref{eq:supp-shape-covariant-transformation}
leaves $\sqrt{|g|}$, $X$, and $\Box\phi$ unchanged. Arbitrary $G_2$
and $G_3\Box\phi$ are separately invariant, while the quartic term
varies only by the leaf-boundary term in
Eq.~\eqref{eq:supp-shape-intrinsic-variation}.

The earlier geometric and coefficient results therefore overlap
substantially with our action. Our additional results are closure
within three-dimensional Horndeski under regular $X$-dependent
disformal maps, the off-shell leaf-shape identity for arbitrary
lower-order terms, and the nonlinear absence of local bulk modes on
the regular $X>0$, $\Xi\neq0$ branch. Boundary charges and the global
gauge quotient require separate analysis.

\section{Four-dimensional Gauss--Bonnet realizations}
\label{sec:supp-GB-realizations}

The aim of this section is to identify when a boundary face of
four-dimensional Gauss--Bonnet geometry reproduces an action built
from the three-dimensional square-root Horndeski term
\begin{equation}
\mathcal L_\circ\equiv
\sqrt{|X|}R+
\frac{s}{2\sqrt{|X|}}
\left[(\Box\phi)^2-\phi_{\mu\nu}\phi^{\mu\nu}\right],
\qquad s=\operatorname{sgn}X .
\end{equation}
A product geometry yields its completed constant-weight action
$S_\circ^{\rm tot}$ Eq. \eqref{eq:endmatter-completed-Horndeski-action}.
A second construction matches $f(\phi)\mathcal L_\circ$ on
restricted local data by choosing the face's extrinsic curvature.

For a region with piecewise smooth non-null boundary, the
four-dimensional Gauss--Bonnet action can be organized as
\begin{equation}
\begin{gathered}
I_{\rm GB}^{(4)}
={}\int_{\mathcal M_4}{\mathrm d}^4x\,\sqrt{|G|}\,\mathcal E_4
+\sum_{\mathcal F\subset\partial\mathcal M_4}I_{\mathcal F}
+I_{\rm joint},\\
I_{\mathcal F}
={}\int_{\mathcal F}{\mathrm d}^3x\,\sqrt{|g|}\,\mathcal B_{\rm M}
+\sum_{\mathcal E\subset\partial\mathcal F}
 \int_{\mathcal E}{\mathrm d}^2x\,\sqrt{|h|}\,
 \mathcal B_{\rm seg}^{(\mathcal F)},\\
I_{\rm joint}
={}4\sum_{\mathcal E}\int_{\mathcal E}
 {\mathrm d}^2x\,\sqrt{|h|}\,
 \psi_{\mathcal E}\mathcal R_{\mathcal E}.
\end{gathered}
\label{eq:supp-GB-action-organization}
\end{equation}
An overall Gauss--Bonnet coupling is suppressed. Here
$\mathcal E_4=R_{ABCD}R^{ABCD}-4R_{AB}R^{AB}+R^2$.
The Myers density $\mathcal B_{\rm M}$ is the Gauss--Bonnet
counterpart of the Gibbons--Hawking--York term on a smooth face.
The term $\mathcal B_{\rm seg}^{(\mathcal F)}$ belongs to an edge
of that face. At a joint where faces meet, $\psi_{\mathcal E}$ is
their signed angle or rapidity and $\mathcal R_{\mathcal E}$ is the
intrinsic scalar curvature of their two-dimensional intersection
\cite{Myers1987,Cano2018}. We use $I_{\rm face}$ for the
contribution $I_{\mathcal F}$ of a selected face.

The boundary densities needed below can be stated in familiar
extrinsic-curvature variables. For a timelike face with outward
spacelike normal, let $\mathbb K_{ij}$ be its extrinsic curvature,
$\mathbb K=\mathbb K^i{}_i$, and $\mathcal G_{ij}[g]$ the Einstein
tensor of its induced metric. On a spacelike edge, let $\sigma^i$
be the outward unit normal within the face. Define
$Q_{ab}=e_a{}^i e_b{}^jD_i\sigma_j$, the edge's extrinsic
curvature within the face, and
$(\mathbb K_\parallel)_{ab}
=e_a{}^i e_b{}^j\mathbb K_{ij}$, the projection of the face's
extrinsic curvature onto the edge. Then
\begin{gather}
\mathcal B_{\rm M}
={}-8\mathcal G_{ij}[g]\mathbb K^{ij}
-\frac{4}{3}\left(
\mathbb K^3
-3\mathbb K\,\mathbb K_{ij}\mathbb K^{ij}
+2\mathbb K^i{}_j\mathbb K^j{}_k\mathbb K^k{}_i
\right),
\nonumber\\
\mathcal B_{\rm seg}^{(\mathcal F)}
=-4\left[
(\mathbb K_\parallel)^a{}_a Q^b{}_b
-(\mathbb K_\parallel)^a{}_b Q^b{}_a
\right].
\label{eq:supp-GB-boundary-densities}
\end{gather}

\subsection{Product geometry and natural Euler descent}
\label{sec:supp-GB-natural-descent}

Consider the product
\begin{equation}
\mathcal M_4=\Sigma\times P_2,\qquad
\mathcal M_3=\Sigma\times C,\qquad
C\subset\partial P_2 .
\label{eq:supp-GB-product-geometry}
\end{equation}
For this explicit realization, equip $\mathcal M_4$ with the product metric. The surface $\Sigma$ is closed, oriented, and Riemannian;
$P_2$ is Lorentzian. Take $C$ to be a smooth timelike segment
with proper time $\tau$ and signed extrinsic curvature $k$,
defined using its outward spacelike normal. Each scalar leaf
$\Sigma_\phi$ of the face $\mathcal M_3$ is a copy of $\Sigma$.

The product has no mixed curvature components, so its Euler
density factorizes. The face extrinsic curvature has only one
nonzero eigenvalue, $k$: its cubic Myers term vanishes, while
the Einstein-tensor term gives $4k{\mathcal R}_{\Sigma_{\phi}}$. The endpoint copies
of $\Sigma$ have $Q_{ab}=0$. Thus
\begin{equation}
\mathcal E_4=2{\mathcal R}_{\Sigma_{\phi}}\mathcal R_P,\qquad
\mathcal B_{\rm M}=4k{\mathcal R}_{\Sigma_{\phi}},\qquad
\mathcal B_{\rm seg}=0,
\label{eq:supp-GB-product-densities}
\end{equation}
where $\mathcal R_P$ is the scalar curvature of $P_2$.

On a patch with $k>0$, take ${\mathrm d}\tau>0$ and define the scalar
locally through the turning of the normal to $C$:
\begin{equation}
{\mathrm d}\phi=4\sqrt2\,k\,{\mathrm d}\tau .
\label{eq:supp-GB-natural-descent-data}
\end{equation}
The induced face metric is $g=-{\mathrm d}\tau^2+h_\Sigma$, giving
$\sqrt X=4k$. Its leaves have $K_{ab}=a_\mu=0$, so the current
$J^\mu$ in
$\mathcal L_\circ=\sqrt X\,{\mathcal R}_{\Sigma_{\phi}}+\nabla_\mu J^\mu$ vanishes.
The selected face contribution therefore reduces to
\begin{align}
I_{\rm face}
&=4\int_C k\,{\mathrm d}\tau
  \int_\Sigma{\mathrm d}^2x\,\sqrt h\,{\mathcal R}_{\Sigma_{\phi}} \nonumber\\
&=\frac{1}{\sqrt2}\int{\mathrm d}\phi
  \int_{\Sigma_\phi}{\mathrm d}^2x\,\sqrt h\,{\mathcal R}_{\Sigma_{\phi}} .
\label{eq:supp-GB-product-normalization}
\end{align}
Because the leaves are closed, their two-dimensional boundary
terms vanish. The completed constant-weight action is
\begin{equation}
S_\circ^{\rm tot}
=\frac{1}{\sqrt2}\int{\mathrm d}\phi
 \int_{\Sigma_\phi}{\mathrm d}^2x\,\sqrt h\,{\mathcal R}_{\Sigma_{\phi}}
=\frac{4\pi\chi(\Sigma)}{\sqrt2}\int{\mathrm d}\phi .
\label{eq:supp-GB-product-leaf-action}
\end{equation}
Hence, on this product geometry,
\begin{equation}
\boxed{I_{\rm face}=S_\circ^{\rm tot}.}
\label{eq:supp-GB-natural-completion}
\end{equation}

The equality is for the selected face. The four-dimensional bulk,
other faces, and genuine joints in
Eq.~\eqref{eq:supp-GB-action-organization} remain separate.
For example, a joint at an endpoint of $C$ contributes
$4\psi E_\Sigma$, where
$E_\Sigma=\int_\Sigma{\mathrm d}^2x\sqrt h\,{\mathcal R}_{\Sigma_{\phi}}
=4\pi\chi(\Sigma)$; its angle or rapidity $\psi$ also depends
on the adjacent face. Keeping the signed normalization of
Eq.~\eqref{eq:supp-GB-natural-descent-data} on a $k<0$ patch
reverses the face equality because $\sqrt X=4|k|$.
Leaves with boundaries require additional side-face and joint
terms.

Separately, a scalar reparametrization gives a covariant
weighted-Euler family. Let $F(\phi)$ be regular, with
$p(\phi)=F_{,\phi}>0$. On a fixed-sign $X\neq0$ patch, direct
substitution into the square-root quartic density gives the
off-shell identity
\begin{align}
\mathcal L_\circ[F(\phi)]
={}&p\,\mathcal L_\circ[\phi]
-4p_{,\phi}\sqrt{|X|}\,\Box\phi
+4Xp_{,\phi\phi}\sqrt{|X|} \nonumber\\
&+\nabla_\mu\!\left(
  2p_{,\phi}\sqrt{|X|}\,\phi^\mu\right).
\label{eq:supp-GB-scalar-pullback}
\end{align}
Setting $f=p$ yields
\begin{equation}
G_4=f(\phi)\sqrt{|X|},\qquad
G_3=-4f_{,\phi}\sqrt{|X|},\qquad
G_2=4Xf_{,\phi\phi}\sqrt{|X|},\qquad
\Xi=0 .
\label{eq:supp-GB-natural-weighted-Euler}
\end{equation}
This identity establishes the correlated coefficients off shell.
Its divergence must be retained when comparing boundary actions.
Regular reparametrizations cover nonvanishing weights locally,
up to an overall sign of the seed action.

\subsection{Restricted realization of the $\Xi\neq0$ action}
\label{sec:supp-GB-Xi-nonzero}

The second construction matches the pure square-root bulk action
\begin{equation}
S[f]=\int_{\mathcal M_3}{\mathrm d}^3x\,\sqrt{|g|}\,
f(\phi)\mathcal L_\circ,\qquad
\Xi=2f_{,\phi},
\label{eq:supp-GB-Sf}
\end{equation}
on patches with $f_{,\phi}\neq0$. It requires a chosen embedding
of the three-dimensional face and restrictions on its scalar
foliation.

Place the timelike face $\mathcal M_3$ at $y=0$ in a local
four-dimensional collar,
\begin{equation}
{\mathrm d} s_4^2={\mathrm d} y^2
+g_{\mu\nu}(y,x){\mathrm d} x^\mu{\mathrm d} x^\nu ,
\label{eq:supp-GB-product-neighborhood}
\end{equation}
and take the region $y\leq0$, so $r=\partial_y$ is its outward
spacelike normal. Within the face, let $n^\mu$ be the timelike
unit normal to the scalar leaves and impose
\begin{equation}
X>0,\qquad a_\mu=0,\qquad
K_{ab}=\frac12K h_{ab}.
\label{eq:supp-GB-restricted-foliation}
\end{equation}
Thus the leaf normal is geodesic and $K_{ab}$ has no shear.
This $K_{ab}$ describes a scalar leaf within the face. The
face's extrinsic curvature in four dimensions is the distinct
tensor $\mathbb K_{ij}$, which we choose as
\begin{equation}
\mathbb K^i{}_j=-\kappa n^i n_j+\ell h^i{}_j .
\label{eq:supp-GB-K-ansatz}
\end{equation}

Assume that the face ends on scalar-aligned spacelike edges, with outward unit normals $\sigma^\mu=\eta n^\mu$, $\sigma_\mu\sigma^\mu=-1$, and
$\eta=\pm1$. Then $Q_{ab}=\eta K_{ab}$ and
$(\mathbb K_\parallel)_{ab}=\ell h_{ab}$, so
Eq.~\eqref{eq:supp-GB-boundary-densities} gives
$\mathcal B_{\rm seg}=-4\eta\ell K$.
The foliation current is $J^\mu=2\sqrt X\,K n^\mu$.
Choosing
\begin{equation}
\ell=-\frac12f(\phi)\sqrt X
\label{eq:supp-GB-ell}
\end{equation}
makes the edge term the outward flux of $fJ^\mu$:
\begin{equation}
\mathcal B_{\rm seg}
=2\eta f\sqrt X\,K
=-\sigma_\mu fJ^\mu .
\label{eq:supp-GB-segment-flux}
\end{equation}

For the same geometry, the Myers density reduces to
\begin{equation}
\mathcal B_{\rm M}
=4\left[
\kappa\left({\mathcal R}_{\Sigma_{\phi}}+\frac12K^2-2\ell^2\right)
+2\ell\,n^\mu\nabla_\mu K+\ell K^2
\right].
\label{eq:supp-GB-Myers-reduced}
\end{equation}
It is linear in $\kappa$. Define
$D={\mathcal R}_{\Sigma_{\phi}}+\frac12K^2-\frac12f^2X$. Where $D\neq0$, choose
\begin{equation}
\kappa=
\frac{
\frac14\bigl[
f\sqrt X\,{\mathcal R}_{\Sigma_{\phi}}-f_{,\phi}\phi_\mu J^\mu
\bigr]
+f\sqrt X\,n^\mu\nabla_\mu K
+\frac12f\sqrt X\,K^2
}{D}.
\label{eq:supp-GB-kappa}
\end{equation}
Together with Eq.~\eqref{eq:supp-GB-ell}, this gives
\begin{equation}
\mathcal B_{\rm M}
=f\sqrt X\,{\mathcal R}_{\Sigma_{\phi}}-f_{,\phi}\phi_\mu J^\mu .
\label{eq:supp-GB-Myers-target}
\end{equation}
Adding the divergence associated with the edge flux now
reproduces the Gauss--Codazzi decomposition:
\begin{equation}
\mathcal B_{\rm M}+\nabla_\mu(fJ^\mu)
=f\sqrt X\,{\mathcal R}_{\Sigma_{\phi}}+f\nabla_\mu J^\mu
=f\mathcal L_\circ .
\label{eq:supp-GB-face-reconstruction}
\end{equation}
Integration over the selected face yields
\begin{equation}
\boxed{I_{\rm face}=S[f]\quad
\text{on the restricted sector}.}
\label{eq:supp-GB-face-action}
\end{equation}

The prescribed face curvature can be realized locally. Extending
the metric as
\begin{equation}
g_{ij}(y,x)=g_{ij}(0,x)+2y\,\mathbb K_{ij}(x)+O(y^2)
\end{equation}
gives the required first normal derivative and preserves the
signature sufficiently near $y=0$. For $f(\phi)=\phi$, the
matched bulk theory has $G_4=\phi\sqrt X$, $G_3=G_2=0$, and
$\Xi=2$.

This match requires Eq.~\eqref{eq:supp-GB-restricted-foliation},
scalar-aligned edges, and $D\neq0$. The coefficient $\kappa$ is
chosen to match the target density; other densities could be
matched by other choices. It therefore gives a restricted local
realization of $S[f]$. Here $S[f]$ is the three-dimensional bulk
action: its independent Dirichlet completion and the genuine
four-dimensional joints remain separate.

\end{document}